\documentclass[aps,12pt,showpacs,showkeys]{revtex4}

\usepackage[english]{babel}
\usepackage{amsmath}
\usepackage{amssymb}
\usepackage{amsbsy}
\usepackage{amstext}
\usepackage{graphicx}
\usepackage{pst-node}
\usepackage{verbatim}

\newcommand{\be}{\begin{eqnarray}}
\newcommand{\ee}{\end{eqnarray}}
\newcommand{\bdm}{\begin{displaymath}}
\newcommand{\edm}{\end{displaymath}}
\newcommand{\ds}{\displaystyle}
\newcommand{\ba}{\begin{array}}
\newcommand{\ea}{\end{array}}
\newcommand{\pa}[1]{\left(#1\right)}
\newcommand{\paq}[1]{\left[#1\right]}
\newcommand{\pag}[1]{\left\{#1\right\}}

\newcommand{\dpa}{\partial}

\newcommand{\ccdot}{.}

\begin{document}

\title{The dynamics of the gravitational two-body problem at fourth 
  post-Newtonian order and at quadratic order in the Newton constant}

\author{Stefano Foffa$^{\rm 1}$ and Riccardo Sturani$^{\rm 2,3}$}

\affiliation{$(1)$ D\'epartement de Physique Th\'eorique and Centre for Astroparticle Physics, Universit\'e de 
             Gen\`eve, CH-1211 Geneva, Switzerland\\
             $(2)$ Dipartimento di Scienze di Base e Fondamenti, 
             Universit\`a di Urbino, I-61029 Urbino, Italy\\
             $(3)$ INFN, Sezione di Firenze, I-50019 Sesto Fiorentino, Italy}

\email{stefano.foffa@unige.ch, riccardo.sturani@uniurb.it}

\begin{abstract}
We derive the conservative part of the Lagrangian and the energy of a 
gravitationally bound two-body system at fourth post-Newtonian order, up to 
terms quadratic in the Newton constant. We also show that such terms are 
compatible with Lorentz invariance and we write an ansatz for the center-of 
mass position.
The remaining terms carrying higher powers of the Newton constant are currently
under investigation.
\end{abstract}



\maketitle

\section{Introduction}

The post-Newtonian approximation to the 2-body problem in General relativity
represents the common approach to study the bound object dynamics in the 
weak curvature, slow motion regime, see \cite{Blanchet_living} and 
\cite{Futamase} for reviews.
The interest in detailed analytical study of the gravitationally bound two-body 
problem has been revived by several concurrent factors.

On the experimental side the era of gravitational wave astronomy is expected
to start in a few years by with the advent of the advanced gravitational wave
detectors LIGO and Virgo \cite{LIGOVirgo}. Among other signals, these detectors 
will be sensitive to gravitational waves emitted by coalescing binaries, making
urgent to derive the equation of motion of binary black holes with
the highest possible accuracy.
The knowledge of accurate template waveform is needed
for matched filtering techniques, commonly used in data analysis, whose
output is particularly sensitive to the time varying phase of the signal which 
must be computed with $O(1)$ precision \cite{Cutler:1992tc}.

On the theoretical side, since few years now stable numerical-relativity 
(NR) waveforms emitted in the last $O(10)$ orbits of a binary 
system (plus merger and ring-down to a final Kerr black hole)
 are available, see e.g. \cite{lrrNR} for a recent review. As NR waveforms
cannot be extended too early in the \emph{inspiral} phase, during which
binary constituents are far apart and non-relativistic, it appears natural to 
combine PN and NR results to construct complete waveform models: the highest
possible accuracy on the analytical PN side is required in order to
reduce the number of cycles of the NR waveforms needed to build complete, 
reliable waveforms \cite{Ohme:2011zm}.

On the theoretical analytical side, before the present paper, the Hamiltonian 
ruling the conservative dynamics of a gravitationally bound, spin-less binary 
system, has been computed at third PN order in \cite{energy_at_3PN} in the 
Arnowitt-Deser-Misner (ADM) coordinates, calculation later confirmed by 
\cite{energy_at_3PNHarm,deAndrade:2000gf,Blanchet:2003gy} in harmonic 
coordinates, and by \cite{energy_at_3PN_conf}, where a different resolution of 
the source singularity has been adopted.
In \cite{Ledvinka:2008tk} the Hamiltonian formalism in ADM coordinates has
been exploited to derive the Hamiltonian in at first order in $G_N$ and
at all orders in momenta. 

The Lagrangian of the gravitationally bound two body system has been recently re-derived at third PN order in 
\cite{Foffa:2011ub}, with the use of an algorithm (according to a strategy first proposed in \cite{Chu:2008xm}) automatizing
the decomposition of the problem into the sum of terms associated to
Feynman diagrams according to the effective field theory 
(EFT) approach developed in \cite{Goldberger:2004jt}, see \cite{EFTReview} for a review. 
The application of the EFT methods to the PN approximation
of General Relativity has also made possible the derivation of new results, mainly for spinning system,
in both the conservative \cite{EFT-E-spin} and dissipative 
dynamics \cite{EFT-F-spin}. 

Following on our previous work \cite{Foffa:2011ub}, the main result of this 
paper is the computation of the two-body effective action for the conservative
dynamics at 4PN order and up to terms quadratic in Newton's constant $G_N$,
and thus represents a decisive step towards the computation of the full
Hamiltonian dynamics at 4PN order. Using the virial relation $v^2\sim G_NM/r$,
being $r(v)$ the relative distance distance (velocity) of the binary 
constituents with $M$ the total mass, the terms contributing to the
4PN order can be parametrized as $G_N^{5-n}v^{2n}$ with $0\leq n\leq 5$.

The remaining terms at 4PN not presented here involve 569 Feynman graphs whose
computation requires the evaluation of integrals among which the most 
complicated ones are few integrals analogous to 4-loop momentum integral in 
quantum field theory.
The computation of the full Lagrangian at 4PN order is well under way and will
be the subject of a future publication.
We also note that the 4PN conservative contribution of the radiation-reaction 
force has been computed in \cite{tail} and re-derived within EFT methods
in \cite{Foffa:2011bq}.

The paper is organized as follows. In sec.~\ref{main} the fundamental gravity
Lagrangian relevant for the two-body dynamics in the PN approximation is laid
down and in sec.~\ref{method} we give a brief overview of the effective field 
theory methods for gravity, which are applied in the rest of the paper to the 
evaluation of the relevant Feynman graphs giving contribution to the 4PN order 
up $G_N^2$. 
In sec.~\ref{se:linearAcc} standard techniques are applied to get rid of terms 
quadratic in the accelerations, in order to obtain a Lagrangian with terms at 
most linear in the acceleration.
In sec.~\ref{se:energy} the energy is derived and in sec.~\ref{se:lorentz}
the Lorentz invariance is checked for and the center-of-mass position
of the two-body system derived, always at $O(G^2)$ order.
We summarize and conclude in sec.~\ref{se:conclusion}.

\section{Short-scale Lagrangian}
\label{main}
We perform our calculation in the EFT framework along the lines of \cite{Foffa:2011ub}; in this section we resume our notations.
The starting point is the action
\be
\label{action}
S = S_{EH}+S_{GF}+S_{pp}\,,
\ee
the first and third terms being, respectively,
the usual Einstein-Hilbert action\footnote{
We adopt the ``mostly plus'' convention
$\eta_{\mu\nu}\equiv {\rm diag}(-,+,+,+)$, and the Riemann and Ricci tensors are
defined as $R^\mu_{\nu\rho\sigma}=\dpa_\rho\Gamma^\mu_{\nu\sigma}+
\Gamma^\mu_{\alpha\rho}\Gamma^\alpha_{\nu\sigma}-\rho\leftrightarrow\sigma$, 
$R_{\mu\nu}\equiv R^\alpha_{\mu\alpha\nu}$. }
and the world-line point particle action
\be
\label{az_EH}
S_{EH}= 2 \Lambda^2\int {\rm d}^{d+1}x\sqrt{-g}\ R(g)\,,\quad
\!\!\!S_{pp}=-\!\!\sum_{i=1,2} m_i\int {\rm d}\tau_i = 
-\!\!\sum_{i=1,2} m_i\int \sqrt{-g_{\mu\nu}(x^\mu_i) {\rm d}x_i^\mu {\rm d}x_i^\nu}\,,
\ee
with
\be
\Lambda^{-2}\equiv 32 \pi G_N L^{d-3}\,.
\ee
As dimensional regularization will be needed in the computation, $L$ is a 
reference length scale which must cancel out from physical results and $G_N$
is the standard 3+1-dimensional Newton constant.
As to the gauge fixing term $S_{GF}$, we follow \cite{Gilmore:2008gq}:
\be
S_{GF}=- \Lambda^2\int {\rm d}^{d+1}x \sqrt{-g}\,\Gamma_\mu\Gamma^\mu\,,
\ee
with $\Gamma^\mu\equiv \Gamma^\mu_{\alpha\beta}g^{\alpha\beta}$, which
corresponds to the same harmonic gauge adopted in \cite{Blanchet_living}.
Still following \cite{Gilmore:2008gq},  we adopt the standard Kaluza-Klein (KK) 
parametrization  of the metric \cite{NR fields} (a somehow similar 
parametrization was first applied within the framework of a PN calculation in
\cite{annales}):
\be
\label{met_nr}
g_{\mu\nu}=e^{2\phi/\Lambda}\pa{
\ba{cc}
-1 & A_j/\Lambda \\
A_i/\Lambda &\quad e^{- c_d\phi/\Lambda}\gamma_{ij}-
A_iA_j/\Lambda^2\\
\ea
}\,,
\ee
with $\gamma_{ij}=\delta_{ij}+\sigma_{ij}/\Lambda$,
$c_d=2\frac{(d-1)}{(d-2)}$ and $i,j$ running over the $d$ spatial dimensions.
In terms of the metric parametrization (\ref{met_nr}),
each world-line coupling to the gravitational degrees of freedom
$\phi$, $A_i$, $\sigma_{ij}$  reads
\renewcommand{\arraystretch}{1.4}
\be
\label{matter_grav}
S_{pp}=-m\ds \int {\rm d}\tau = \ds-m\int {\rm d}t\ e^{\phi/\Lambda}
\sqrt{\pa{1-\frac{A_i}{\Lambda}v^i}^2
-e^{-c_d \phi/\Lambda}\pa{v^2+\frac{\sigma_{ij}}{\Lambda}v^iv^j}}\,,
\ee
\renewcommand{\arraystretch}{1.4}
and its Taylor expansion provides the various particle-gravity vertices of the EFT.

Also the pure gravity sector $S_{bulk}=S_{EH}+ S_{GF}$ can be explicitly written
in terms of the KK variables; we report here only those terms which are needed
for the full 4PN calculation (only a part of which is performed in the present work):
\renewcommand{\arraystretch}{1.4}
\be
\label{bulk_action}
\ba{rcl}
\ds S_{bulk} &\supset &\ds \int {\rm d}^{d+1}x\sqrt{-\gamma}
\left\{\frac{1}{4}\left[(\vec{\nabla}\sigma)^2-2(\vec{\nabla}\sigma_{ij})^2-\left(\dot{\sigma}^2-2(\dot{\sigma}_{ij})^2\right){\rm e}^{\frac{-c_d \phi}{\Lambda}}\right]- c_d \left[(\vec{\nabla}\phi)^2-\dot{\phi}^2 {\rm e}^{-\frac{c_d\phi}{\Lambda}}\right]\right.\\
&&\ds
+\left[\frac{F_{ij}^2}{2}+\left(\vec{\nabla}\!\!\cdot\!\!\vec{A}\right)^2 -\dot{\vec{A}}^2 {\rm e}^{-\frac{c_d\phi}{\Lambda}} \right]
{\rm e}^{\frac{c_d \phi}{\Lambda}}+\frac 2\Lambda\paq{\pa{F_{ij}A^i\dot{A^j}+\vec{A}\!\!\cdot\!\!\dot{\vec{A}}(\vec{\nabla}\!\!\cdot\!\!\vec{A})}
{\rm e}^{\frac{c_d \phi}{\Lambda}}-c_d\dot{\phi}\vec{A}\!\!\cdot\!\!\vec{\nabla}\phi}
\\
&&\ds
+2 c_d \left(\dot{\phi}\vec{\nabla}\!\!\cdot\!\!\vec{A}-\dot{\vec{A}}\!\!\cdot\!\!\vec{\nabla}\phi\right)
+\frac{\dot{\sigma}_{ij}}{\Lambda}\left(-\delta^{ij}A_l\hat{\Gamma}^l_{kk}+ 2A_k\hat{\Gamma}^k_{ij}-2A^i\hat{\Gamma}^j_{kk}\right)-c_d\frac{\dot{\phi}^2\vec{A}^2}{\Lambda^2}\\
&&\ds
-\left.\frac{1}{\Lambda}\left(\frac{\sigma}{2}\delta^{ij}-\sigma^{ij}\right)
\left({\sigma_{ik}}^{,l}{\sigma_{jl}}^{,k}-{\sigma_{ik}}^{,k}{\sigma_{jl}}^{,l}+\sigma_{,i}{\sigma_{jk}}^{,k}-\sigma_{ik,j}\sigma^{,k}
\right)\right\}\,.
\ea
\ee
\renewcommand{\arraystretch}{1.}
The form (\ref{bulk_action}) of the gravitational action
\footnote{$\hat{\Gamma}^i_{jk}$ is the connection of the purely spatial metric $\gamma_{ij}$
, $F_{ij}\equiv A_{j,i}- A_{i,j}$ and indices must be raised and
contracted via the $d$-dimensional metric tensor $\gamma$,
although in the last three lines of eq.~\ref{bulk_action} one can use just 
$\delta_{ij}$ as neglected terms are not needed at 4PN order; on the other hand
all the spatial derivatives are meant to be simple (not covariant) ones and,
when ambiguities might raise, gradients are always meant to act on
contravariant fields (so that, for instance,
$\vec{\nabla}\!\!\cdot\!\!\vec{A}\equiv\gamma^{ij}A_{i,j}$
and $F_{ij}^2\equiv\gamma^{ik}\gamma^{jl}F_{ij}F_{kl}$).} is in agreement with the one derived in \cite{Kol:2010si}.

\section{Integrating out gravity}
\label{method}
We have now all the ingredients to obtain a 2-body effective action
$S_{eff}$ with manifest power counting in $G_N$ and $v$ at the desired (fourth)
post-Newtonian order.
This can be done by integrating out the graviton fields from the full
action derived above
\begin{equation}
\label{eq:seff}
  S_{eff}=\int D\phi D\sigma_{ij} DA_k\exp[i(S_{EH}+S_{GF}+ iS_{pp})]\,,
\end{equation}
to obtain an effective action $S_{eff}$ in terms of the particle positions and
their derivatives.
As usual in field theory, the functional integration can be perturbatively 
expanded in terms of Feynman diagrams involving the gravitational 
degrees of freedom as internal lines
\footnote{As we focus on the conservative part of the dynamics, internal lines
in Feynman diagrams correspond to the so-called \emph{potential} gravitons and
no gravitational radiation is emitted.}, 
regarded as dynamical fields emitted and absorbed by the point particles which 
are taken as non-dynamical sources.

In order to allow manifest $v$  scaling it is convenient to work with the 
space-Fourier transformed fields
\renewcommand{\arraystretch}{1.4}
\be
\label{Fourk}
W^a_{\bf p}(t)  \equiv \ds\int {\rm d}^dx\, W^a(t,x) e^{-i{\bf p}\cdot x}\,\quad
 {\rm with\ } W^a=\{\phi,A_i,\sigma_{ij}\}\,.
\ee
\renewcommand{\arraystretch}{1}
The fields defined above are the fundamental variables in terms of
which we are going to construct the Feynman graphs; the action governing
their dynamics can be found from eqs.~(\ref{matter_grav},\ref{bulk_action}).

By looking in particular at the quadratic parts, one can explicitly write the 
propagators:
\be
\label{propagators}
\ds P[W^a_{\bf p}(t_a)W^b_{\bf p'}(t_b)]&=&\ds \frac{1}{2} P^{aa}\delta_{ab}
\ds (2\pi)^d\delta^{d}({\bf p}+{\bf p'}){\cal P}({\bf p}^2,t_a,t_b)\delta(t_a-t_b)\,,
\ee
where $P^{\phi\phi}=-\frac{1}{c_d}$, $P^{A_iA_j}=\delta_{ij}$,
$P^{\sigma_{ij}\sigma_{kl}}=-\left(\delta_{ik}\delta_{jl}+\delta_{il}\delta_{jk}+(2-c_d)\delta_{ij}\delta_{kl}\right)$ 
and
\be
{\cal P}({\bf p}^2,t_a,t_b)=\frac{i}{{\bf p}^2-\partial_{t_a}\partial_{t_b}}\simeq
\frac{i}{{\bf p}^2}\left(1+\frac{\partial_{t_a}\partial_{t_b}}{{\bf p}^2}+
\frac{\partial_{t_a}^2\partial_{t_b}^2}{{\bf p}^4}\dots\right)
\ee
is the full relativistic propagator,
which have been expanded as an instantaneous non-relativistic
part plus insertion terms involving time derivatives (which after hitting the 
$e^{i{\bf p}\cdot x}$ factors bring in extra $v$ factors) and
the infinite series is truncated at the some finite order in every Feynman 
diagram.

The next step is to to lay down and compute all the relevant Feynman diagrams;
this has been done by means of a new version of the Mathematica code 
\cite{Mathematica}, including the use of the FeynCalc software 
\cite{Mertig:1990an}, which has been used to reproduce the 3PN Lagrangian
\cite{Foffa:2011ub}.

\subsection{Order $G_N$}
\begin{figure}
\includegraphics[width=1.\linewidth]{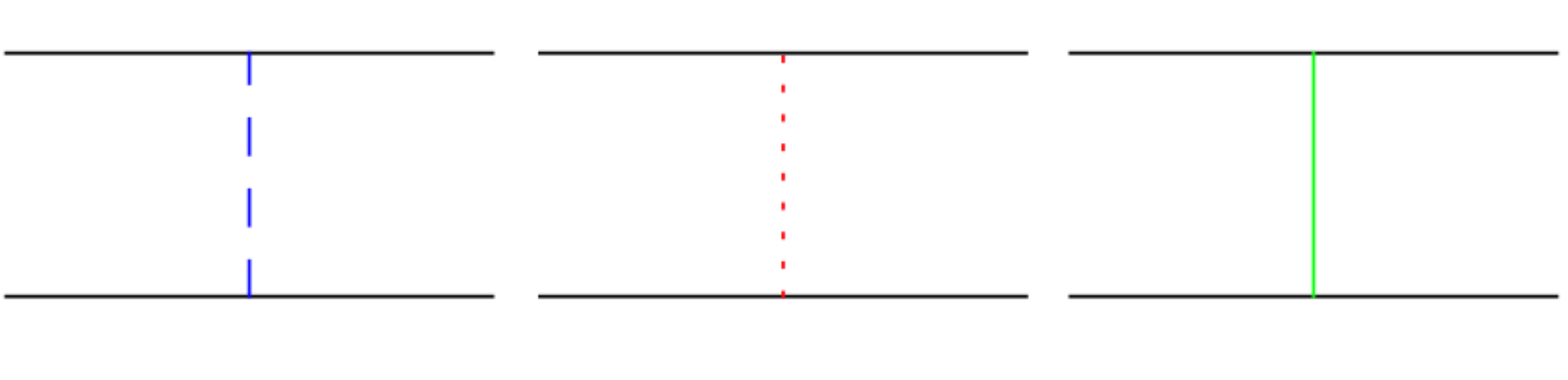}
\caption{The three diagrams contributing at order $G_N$.  The $\phi$, $A$ and 
  $\sigma$ propagators are represented respectively by blue dashed, red dotted 
  and green solid lines. Propagators and vertices must be considered
  at appropriate $v$ order to pick the 4PN contribution of these 
  graphs.}
\label{diaG1}
\end{figure}

The three relevant diagrams at this order are shown in fig.~\ref{diaG1}, the 
typical graph contribution $A$ to the effective action (\ref{eq:seff}) has the 
following form:
\be
A \sim G_N m_1 m_2 L^{d-3}\int_{{\bf p},t_1,t_2}\sum_n
\frac{(\partial_{t_1}\partial_{t_2})^n}{{\bf p}^{2(n+1)}}
{\rm e}^{i {\bf p}\cdot(x_1- x_2)}V_1V_2\,,
\ee
where integration is performed over times $t_{1,2}$ and the measure of momentum 
integration is ${\rm d}^dp/(2\pi)^d$, $n$ is an 
integer number with $n\leq 4$, $x_{1,2}\equiv x_{1,2}(t_{1,2})$, and 
$V_{1,2}\equiv V_{1,2}[v_{1,2}(t_{1,2})]$ are the appropriate point particle 
vertices, expanded at the appropriate order in $v_{1,2}$, which can be read from
eq.~(\ref{matter_grav}).
The integrations can be performed via standard Fourier formulae and do 
not lead to any divergence in dimensional regularization.

We report below the result of the sum of the diagrams
(the value of every single diagram of this paper is available on a Mathematica notebook at \cite{nbrelease}); symmetrization under 
exchange of particles $+ (1\leftrightarrow 2)$ is understood throughout all 
this paper
\be\label{G1tot}
{\cal L}^{4PN}_{G_N}&=&\frac{G_N m_1 m_2}{r}\left\{v_2\ccdot a_1\left[\frac{3}{16}\left(v_1\ccdot v_2\right)^2 v_1^r -v_2^4\left(\frac{49}{64}v_1^r+\frac{123}{64}v_2^r\right)-\frac{33}{32}v_1^2 v_2^2v_2^r\right. \right. \nonumber\\
&+&v_2^2 v_1\ccdot v_2\left(\frac{1}{16}v_1^r+\frac{3}{16}v_2^r\right)+v_2^2\left(\frac{17}{96} {v_1^r}^3+\frac{7}{32}{v_1^r}^2v_2^r+\frac{13}{32}v_1^r{v_2^r}^2+\frac{77}{96} {v_2^r}^3\right) \nonumber\\
&+&\left.v_1\ccdot v_2\left(\frac{1}{16} {v_1^r}^2+\frac{5}{32}v_1^rv_2^r+\frac{5}{32}{v_2^r}^2\right)v_1^r-{v_1^r}^3\left(\frac{13}{64}{v_1^r}^2+\frac{11}{64} v_1^r v_2^r+\frac{5}{32} {v_2^r}^2\right)\right]r\nonumber\\
&+&v_1\ccdot a_1 v_2^2\left[v_1^2\left(\frac{123}{32} v_1^r+\frac{93}{32}v_2^r\right)-\frac{33}{16}v_1\ccdot v_2v_2^r- {v_2^r}^2\left(\frac{23}{32}v_1^r+\frac{77}{96}v_2^r\right)\right]r\nonumber\\
&+& r a_1^r\left[\frac{123}{128} v_1^4 v_2^2+ \frac{1}{16} \left(v_1\ccdot v_2\right)^3 + \frac{1}{32} v_2^2 \left(v_1\ccdot v_2\right)^2-\frac{49}{64} v_2^4 v_1\ccdot v_2+\frac{75}{128} v_2^6-\frac{23}{64}v_1^2v_2^2 {v_2^r}^2\right.\nonumber\\
&+&\left(v_1\ccdot v_2\right)^2\left(\frac{3}{32} {v_1^r}^2+\frac{5}{32} v_1^r v_2^r +\frac{5}{64} {v_2^r}^2\right)+v_2^2 v_1\ccdot v_2\left(\frac{17}{32} {v_1^r}^2+\frac{7}{16} v_1^r v_2^r +\frac{13}{32} {v_2^r}^2\right)\nonumber\\
&-&v_2^4\left(\frac{57}{128} {v_1^r}^2+\frac{31}{64} v_1^r v_2^r +\frac{53}{128} {v_2^r}^2\right)-v_1\ccdot v_2\left(\frac{65}{64} {v_1^r}^2+\frac{11}{16} v_1^r v_2^r +\frac{15}{32} {v_2^r}^2\right){v_1^r}^2\nonumber\\
&+&v_2^2\left(\frac{65}{128} {v_1^r}^4+\frac{15}{32} {v_1^r}^3v_2^r +\frac{27}{64} {v_1^r}^2 {v_2^r}^2 +\frac{11}{32}v_1^r{v_2^r}^3+\frac{27}{128} {v_2^r}^4\right)\nonumber\\
&-&{v_2^r}^4\left.\left(\frac{15}{128} {v_1^r}^2+\frac{5}{64} v_1^r v_2^r +\frac{5}{128} {v_2^r}^2\right)\right]\nonumber\\
&+&\frac{75}{128} v_1^8-\frac{1}{32}\left(v_1\ccdot v_2\right)^4
+\frac{203}{128}v_1^6v_2^2-\frac{5}{4}v_1^6v_1\ccdot v_2+\frac{1}{16}\left(v_1\ccdot v_2\right)^3v_1^2+\frac{3}{256} v_1^4v_2^4\nonumber\\
&+&\frac{3}{4}v_1^4\left(v_1\ccdot v_2\right)^2
-\frac{11}{32}v_1^6{v_2^r}^2-\frac{111}{64}v_1^4v_2^2 v_1\ccdot v_2 -v_1^4v_2^2\left(\frac{57}{128} {v_2^r}^2+\frac{123}{128} {v_1^r}^2-\frac{15}{64}v_1^r v_2^r\right)\nonumber\\
&+&\frac{3}{4}v_1^4v_1\ccdot v_2{v_2^r}^2
+\left(v_1\ccdot v_2\right)^3\left(\frac{1}{32}v_2^r-\frac{5}{64}v_1^r\right)v_1^r\nonumber\\
&+&\left(v_1\ccdot v_2\right)^2v_1^2\left(\frac{3}{32}{v_1^r}^2 +\frac{5}{32}v_1^r v_2^r-\frac{27}{64}{v_2^r}^2\right)
+v_1^2v_1\ccdot v_2 v_2^2\left(\frac{17}{32}v_1^r+\frac{3}{8}v_2^r\right)v_1^r
\nonumber\\
&+&\left(v_1\ccdot v_2\right)^2\left(\frac{3}{4} {v_1^r}^2+\frac{15}{64} v_1^r v_2^r+\frac{15}{64}{v_2^r}^2\right){v_1^r}^2\nonumber\\
&+& \frac{21}{64}v_1^4{v_2^r}^4-v_1^2v_1\ccdot v_2\left(\frac{65}{64} {v_1^r}^4+\frac{11}{16}{v_1^r}^3v_2^r
+\frac{15}{32}{v_1^r}^2{v_2^r}^2+\frac{3}{4} {v_2^r}^4\right)\nonumber\\
&+&v_1^2v_2^2\left(\frac{65}{128}{v_1^r}^2+\frac{15}{32}v_1^r v_2^r+\frac{3}{4}{v_2^r}^2\right){v_1^r}^2
-v_1^2{v_2^r}^4\left(\frac{15}{128}{v_1^r}^2+\frac{5}{64}v_1^rv_2^r+\frac{65}{128}{v_2^r}^2\right)\nonumber\\
&+&\left.v_1\ccdot v_2\left(\frac{65}{64}{v_1^r}^3- \frac{5}{32}{v_1^r}^2v_2^r-\frac{5}{64}v_1^r{v_2^r}^2-\frac{15}{32}{v_2^r}^3\right){v_1^r}^3
+\frac{35}{256} {v_1^r}^4{v_2^r}^4\right\}+{\cal L}^{a^2}_{G_N}\,,
\ee
where $r\equiv |x_1-x_2|$, $v_{1,2}^r\equiv v_{1,2}\ccdot(x_1-x_2)/r$, 
$a_{1,2}^r\equiv a_{1,2}\ccdot(x_1-x_2)/r$, 
and terms depending quadratically on accelerations and their derivatives have 
been isolated in
\be
{\cal L}^{a^2}_{G_N}&=&G_N m_1 m_2 r \left\{\left[\frac{31}{11520}\ddot{a}_1\ccdot \ddot{a}_2-\frac{1}{2304}\ddot{a}_1^r \ddot{a}_2^r\right]r^4
+\left[\frac{1}{96}\ddot{a}_1^r\dot{a}_2^rv_1^r-\frac{7}{288}\ddot{a}_1^r v\ccdot \dot{a}_2-\frac{23}{288}\dot{a}_2\ccdot \ddot{a}_1 v_1^r\right]r^3\right.\nonumber\\
&+&\left[\left(\frac{1}{32}v_1\ccdot v_2-\frac{133}{576}v_1^2-\frac{91}{576}a_1^r r-\frac{91}{192}{v_1^r}^2-\frac{11}{96}v_1^r v_2^r\right)\dot{a}_1\dot{a}_2
+\left(\frac{11}{192}v_1^2-\frac{1}{32}v_1\ccdot v_2+\frac{5}{192}a_1^r r\right.\right.\nonumber\\
&+&\left.\frac{5}{192}{v_1^r}^2+\frac{1}{96}v_1^r v_2^r\right)\dot{a}_1^r\dot{a}_2^r
+\left(\frac{9}{32}v\ccdot \dot{a}_1  v_2^r+\frac{1}{6}v_1\ccdot \dot{a}_1v_1^r-\frac{7}{48}v_2\ccdot \dot{a}_1 v_1^r+\frac{59}{576}a_2\ccdot \dot{a}_1 r+\frac{3}{64}a_1\ccdot \dot{a}_1 r\right)\dot{a}_2^r\nonumber\\
&+&\left.\frac{1}{32}v_2\ccdot \dot{a}_1v_1\ccdot \dot{a}_2 -\frac{2}{9}v_1\ccdot \dot{a}_1 v_1\ccdot \dot{a}_2+\frac{1}{9}v_1\ccdot \dot{a}_1 v_2\ccdot \dot{a}_2\right] r^2
+\left[\left(\frac{59}{64}a_1^r v_1^r r+\frac{29}{64}a_1^r v_2^r r-\frac{19}{192}v_2\ccdot a_1 r\right.\right.\nonumber\\
&+&\left.\frac{37}{64}v_1\ccdot a_1 r+\frac{85}{64}v_1^2v_1^r+\frac{55}{64}v_1^2 v_2^r+\frac{5}{32}v_1\ccdot v_2 v_1^r+\frac{59}{384}({v_1^r}^3+{v_2^r}^3)+\frac{29}{128}(v_1^r + v_2^r)v_1^r v_2^r\right)a_2\ccdot \dot{a}_1\nonumber\\
&+&\left(\frac{37}{64}a_1\ccdot a_2 r
+\frac{67}{384}a_2^2 r+\pag{\frac{27}{64}v_1\ccdot a_2-\frac{145}{64}v_2\ccdot a_1-\frac{59}{64}v_2\ccdot a_2 }v_1^r\right.\nonumber\\
&+&\left.\pag{\frac{13}{64}v_1\ccdot a_2-\frac{27}{64}v_2\ccdot a_1-\frac{35}{64}v_2\ccdot a_2 }v_2^r\right)v_1\ccdot \dot{a}_1+\left(\pag{\frac{37}{64}v_2\ccdot a_1+\frac{53}{64}v_2\ccdot a_2-\frac{145}{64}v_1\ccdot a_1-\frac{3}{64}v_1\ccdot a_2}\right. v_1^r\nonumber\\
&+&\left.\left.\pag{\frac{3}{64}(v_2\ccdot a_1-v_1\ccdot a_2)-\frac{27}{64}v_1\ccdot a_1+\frac{67}{64}v_2\ccdot a_2}v_2^r
-\frac{19}{192}a_1\ccdot a_2 r-\frac{67}{384}a_2^2 r\right)v_2\ccdot \dot{a}_1\right]r\nonumber\\
&+&\left[\left(\left\{\frac{59}{64}a_1\ccdot a_2+\frac{27}{128}a_2^2-\frac{5}{64}a_1^r a_2^r \right\}v_1^r
+\pag{\frac{29}{64}a_1\ccdot a_2+\frac{29}{128}a_2^2-\frac{3}{64}a_1^r a_2^r}v_2^r\right.\right.\nonumber\\
&+&\left.\frac{27}{64}v\ccdot a_2 a_1^r+\frac{17}{64}v_2\ccdot a_2 a_2^r-\frac{13}{64}v_1\ccdot a_2  a_2^r\!\right)\! r\!
+\!\left(\frac{23}{64}v_2^2+\frac{5}{32}v_1\ccdot v_2+\frac{13}{64}{v_2^r}^2+\frac{13}{32}v_1^r v_2^r\!\right)\!v_1\ccdot a_2\nonumber\\
&+&\left(\frac{37}{64}v_1^2+\frac{27}{64}{v_1^r}^2\right)v\ccdot a_2+\left(\frac{9}{32}v_1\ccdot v_2
-\frac{51}{64} v_2^2-\frac{17}{64}{v_2^r}^2-\frac{15}{32}v_1^r v_2^r\right)v_2\ccdot a_2
+\left(\frac{5}{32}v_1\ccdot v_2v_1^r -\frac{11}{64} v_1^2 v_1^r\right.\nonumber\\
&-&\left.\left.\frac{9}{64}v_1^2 v_2^r+\frac{5}{192}{v_1^r}^3+\frac{3}{64} {v_1^r}^2 v_2^r\right)a_2^r\right]r \dot{a}_1^r
+\left[\left(\frac{37}{64}a_1^2-\frac{19}{384}a_1\ccdot a_2+\frac{145}{128}{a_1^r}^2
+\frac{1}{8}a_1^r a_2^r\right)a_1\ccdot a_2\right.\nonumber\\
&+&\left.\frac{67}{768}a_1^2 a_2^2-\frac{3}{256}{a_1^r}^2{a_2^r}^2-\frac{5}{128}{a_1^r}^3 a_2^r\right] r^2
+\left[\left(\frac{27}{64}v_1\ccdot a_2-\frac{145}{64}v_2\ccdot a_1-\frac{59}{64}v_2\ccdot a_2 \right)v_1^r\right.\nonumber\\
&+&\left.\left(\frac{13}{64}v_1\ccdot a_2-\frac{27}{64}v_2\ccdot a_1-\frac{35}{64}v_2\ccdot a_2\right) v_2^r \right] r a_1^2
+\left[\frac{111}{32} v_1\ccdot a_1 v_1^r+\frac{7}{32}v_2^r v_1\ccdot a_1-\frac{17}{32}v_2\ccdot a_1v_1^r\right.\nonumber\\
&-&\left.\frac{5}{32}v_2\ccdot a_1 v_2^r +\left(\frac{181}{64}v_1^2-\frac{5}{32}v_1\ccdot v_2
-\frac{3}{32}v_2^2+\frac{231}{128}{v_1^r}^2+\frac{25}{64}v_1^r v_2^r-\frac{3}{128}{v_2^r}^2\right)a_1^r\right] r a_1\ccdot a_2\nonumber\\
&+&\left[\frac{65}{64}v_1^rv\ccdot a_2+\left(\frac{11}{64}v_1\ccdot a_2-\frac{15}{64}v_2\ccdot a_2\right)v_2^r\right.\nonumber\\
&+&\left.\left(\frac{11}{64}v_1\ccdot v_2-\frac{27}{128}v_1^2+\frac{3}{128}v_2^2+\frac{15}{128}{v_1^r}^2+\frac{5}{64}v_1^r v_2^r-\frac{3}{128}{v_2^r}^2\right)a_2^r\right]r {a_1^r}^2\nonumber\\
&+&\left[\frac{15}{64}v_2\ccdot a_1v_1^r-\frac{27}{64}v_1\ccdot a_1 v_1^r+\frac{3}{64}v_2\ccdot a_1v_2^r-\frac{21}{64}v_1\ccdot a_1 v_2^r \right]r a_1^r a_2^r
+\left(\frac{3}{128}v_2\ccdot a_1-\frac{27}{64} v_1\ccdot a_1\right)v_2\ccdot a_1 r a_2^r\nonumber\\
&+&\left[\left(2 v_1\ccdot a_2-\frac{145}{64} v_2\ccdot a_1-\frac{5}{2}v_2\ccdot a_2\right)v_1\ccdot a_1
+\left(\frac{37}{128}v_2\ccdot a_1-\frac{5}{16}v_1\ccdot a_2+\frac{49}{32}v_2\ccdot a_2\right)v_2\ccdot a_1\right] r a_1^r\nonumber\\
&+&\left[\frac{299}{128}v_1^4-\frac{1}{4}\left(v_1\ccdot v_2\right)^2-\frac{19}{32}v_1^2 v_1\ccdot v_2+\frac{17}{64}v_1^2 v_2^2
+\left(\frac{303}{128}{v_1^r}^2+\frac{45}{64}v_1^r  v_2^r+\frac{17}{128}{v_2^r}^2\right)v_1^2\right.\nonumber
\ee
\be\label{doubzG1}
&+&\left.\frac{5}{64}(v_1^r+v_2^r) v_1\ccdot v_2 v_1^r-\!\left(\!\frac{43}{192}{v_1^r}^2+\frac{5}{24}v_1^r v_2^r+\frac{13}{128}{v_2^r}^2\!\right)\!{v_1^r}^2\!\right]\!a_1\ccdot a_2
+\left[\frac{13}{32}v_1^2 v_1\ccdot v_2-\frac{53}{128}v_1^4-\frac{7}{64}v_1^2 v_2^2\right.\nonumber\\
&+&\left.\!\left(\!\frac{27}{64}{v_1^r}^2+\frac{11}{32}v_1^r  v_2^r+\frac{3}{32}{v_2^r}^2\!\right)\!v_1^2
-\!\left(\!\frac{11}{32}v_1^r+\frac{1}{8}v_2^r\!\right)\!v_1^r v_1\ccdot v_2
-\!\left(\!\frac{15}{128}v_1^r+\frac{5}{32}v_2^r\!\right)\!{v_1^r}^3\!\right]\!a_1^r a_2^r\nonumber\\
&+&\left[\left(\left\{\frac{19}{32}v_1\ccdot v_2-\frac{23}{32}v_2^2-\frac{73}{64}v_1^2\right\} v_1^r
+\left\{\frac{17}{16}v_1\ccdot v_2-\frac{123}{64}v_1^2\right\} v_2^r
+\frac{9}{64}{v_1^r}^3+\frac{21}{64}{v_1^r}^2v_2^r+\frac{15}{32}v_1^r{v_2^r}^2\right)\!v_1\ccdot a_1\right.\nonumber\\
&+&\left.\left(\left\{\frac{25}{64}v_1^2+\frac{5}{32}v_1\ccdot v_2+\frac{13}{32}v_2^2\right\} v_1^r
+\left\{\frac{43}{64}v_1^2+\frac{3}{16}v_1\ccdot v_2\right\} v_2^r
-\frac{5}{64}{v_1^r}^3-\frac{13}{64}{v_1^r}^2v_2^r-\frac{11}{32}v_1^r{v_2^r}^2\right)\!v_2\ccdot a_1\right.\nonumber\\
&+&\left.\left(\frac{65}{96}{v_2^r}^2-\frac{77}{32}v_2^2 \right)v_2^r v\ccdot a_1\right]a_2^r
+\left[\frac{161}{32}v_1^2-\frac{5}{8}v_1\ccdot v_2+\frac{47}{16}v_2^2+2{v_1^r}^2
+\frac{13}{8}v_1^r v_2^r+\frac{49}{32}{v_2^r}^2\right]v_1\ccdot a_1 v_1\ccdot a_2\nonumber\\
&+&\left[\frac{59}{32}(v_1\ccdot v_2+v_1^r v_2^r)-\frac{145}{64}(v_1^2+{v_1^r}^2)+\frac{27}{64}(v_2^2+{v_2^r}^2)\right]v_1\ccdot a_1 v_2\ccdot a_1
+\left[\frac{119}{64}v_1\ccdot v_2-\frac{177}{32}v_1^2-\frac{5}{2}{v_1^r}^2\right.\nonumber\\
&-&\left.\frac{65}{64}v_1^r v_2^r\right]v_1\ccdot a_1 v_2\ccdot a_2
-\left[\frac{29}{32}v_1^2+\frac{27}{64}v_1\ccdot v_2+\frac{5}{16}{v_1^r}^2+\frac{3}{64}v_1^r v_2^r\right]v_2\ccdot a_1 v_1\ccdot a_2\nonumber\\
&+&\left.\left[\frac{37}{128}(v_1^2+{v_1^r}^2)-\frac{17}{64}(v_1\ccdot v_2+v_1^r v_2^r)-\frac{3}{128}(v_2^2+{v_2^r}^2)\right]\left(v_2\ccdot a_1\right)^2\right\}\,.
\ee

The rationale for this separation is that such terms can be further reduced via
the {\it double zero trick} (see e.g. \cite{multizero}) which consists in 
writing, for instance
\renewcommand{\arraystretch}{1.4}
\be
\label{dzero}
\ba{rcl}
\ds a_1\cdot a_2&=&\ds 
\left(a_1-eq_1+eq_1\right)\cdot\left(a_2-eq_2+eq_2\right)\\
&=&\ds\left(a_1-eq_1\right)\cdot\left(a_2-eq_2\right)+eq_1\cdot a_2+a_1\cdot eq_2
-eq_1\cdot eq_2\\
&\equiv&\ds (a_1- eq_1)\cdot(a_2-eq_2)+(a_1\cdot a_2)_{z^2}\,.
\ea
\ee
\renewcommand{\arraystretch}{1.}
being $eq_{1,2}$ the value of $a_{1,2}$ in terms of $r$, $v_1$ and $v_2$ as 
dictated by the equations of motion at the desired PN order.
Since the first term on the last line of eq.~(\ref{dzero}) gives a vanishing 
contribution to the equations of motion, it can be dropped from the Lagrangian, 
thus leaving only terms linear in the accelerations, which however carry higher
powers of $G_N$, as $eq_{1,2}={\cal O}(G_N)$.
The standard form of the 3PN Lagrangian, as it is for instance reported in 
eq.~(174) of \cite{Blanchet_living}, is obtained via reiterated use of this 
trick.

\subsection{Order $G_N^2$}
The 23 diagrams contributing at ${\cal O}(G^2_N)$ are shown in figure 
\ref{diaG2}. These diagrams are either factorizable into two single-graviton 
exchanges, or give a contribution $A$ to the effective action of the type:
\be
A &\sim& G^2_N m_1^2 m_2 L^{2(d-3)}\sum_{n_a,n_b,n_2}\int_{t,t_{1a},t_{1b},t_2}
(i\partial_{t_{1a}})^{2n_a}(i\partial_{t_{1b}})^{2n_b}(i\partial_{t_2})^{2n_2}
V_{1a}V_{1b}V_2
\nonumber\\
&\times&\int_{{\bf p},{\bf q}}\frac{{\rm e}^{i [({\bf p-q})\cdot x_1(t_{1a})+{\bf q}\cdot x_1(t_{1b})-{\bf p}\cdot x_2(t_2)]}}{({\bf p-q})^{2(n_a+1)}{\bf q}^{2(n_b+1)}{\bf p}^{2(n_2+1)}}
V_{bulk}\delta(t-t_{1a})\delta(t-t_{1b})\delta(t-t_2)\,,
\ee
with the $n_i\leq 3$ and $V_{bulk}\equiv V_{bulk}[{\bf p},{\bf q},\partial_{t_{1a}},\partial_{t_{1b}},\partial_{t_2}]$ 
being the bulk vertices describing the appropriate three-graviton interaction.

These integrals can be expressed in terms of standard ones, eventually after 
some integration by parts, leading to the following total result:
\begin{figure}
\includegraphics[width=1.\linewidth]{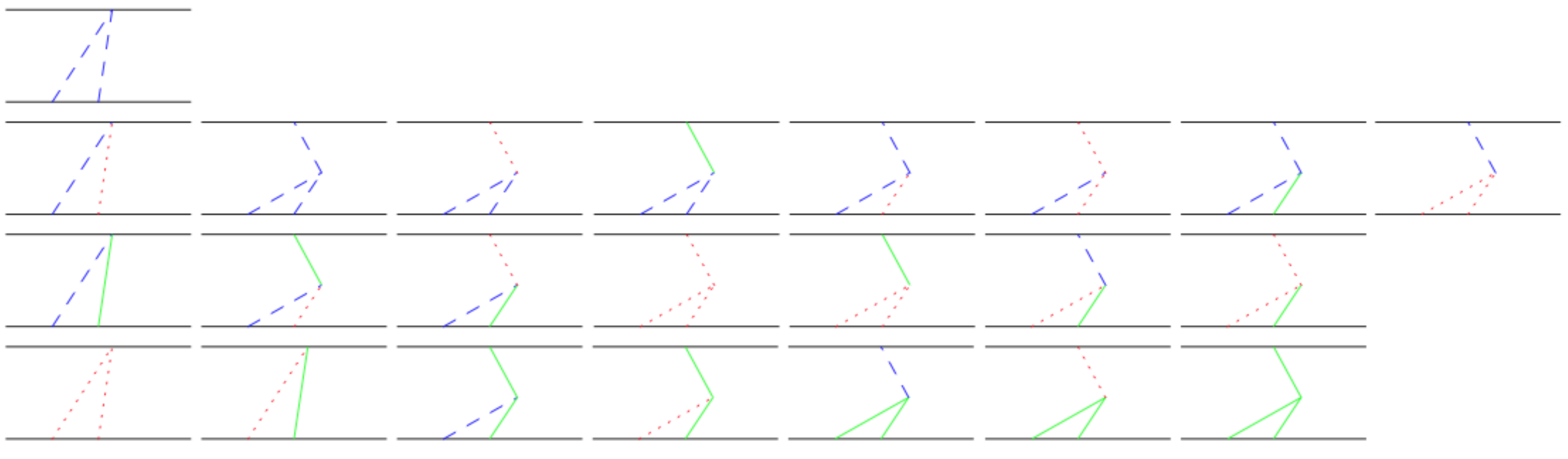}
\caption{The diagrams contributing at order $G^2_N$ arranged in 4 lines. 
Diagrams in the $n^{\rm th}$ line enter the dynamics at $n$PN order.
Propagators and vertices must be considered at appropriate $v$ order to pick 
the 4PN contribution of these diagrams.}
\label{diaG2}
\end{figure}
\be
{\cal L}^{4PN}_{G^2_N}&=&\frac{G_N^2 m_1^2 m_2}{r^2}\left\{v_1\ccdot a_1\left[v_2^2 \left(\frac{28}{15}v_1^r+\frac{133}{10}v_2^r\right)
-v_1\ccdot v_2\left(\frac{56}{15}v_1^r+\frac{314}{15}v_2^r\right)\right]r\right.\nonumber\\
&+&v_2\ccdot a_1\left[v_1^2\left(\frac{47}{15}v_1^r -\frac{157}{15}v_2^r\right)+ v_1.v_2\left(\frac{229}{15}v_2^r-\frac{29}{15} v_1^r\right)\right.\nonumber\\
&+&\left.v_2^2 \left(\frac{41}{20}v_1^r-\frac{133}{10}v_2^r\right)
-\frac{103}{15}{v_1^r}^3+\frac{304}{15}{v_1^r}^2v_2^r-\frac{93}{5}v_1^r{v_2^r}^2+\frac{133}{15}{v_2^r}^3\right]r\nonumber\\
&+&a_1^r\left[\frac{49}{6}v_1^4+\frac{117}{10}\left(v_1\ccdot v_2\right)^2+\frac{469}{80}v_2^4
+\frac{293}{30}v_1^2v_2^2-\frac{293}{15}v_1^2v_1\ccdot v_2-\frac{937}{60}v_1\ccdot v_2v_2^2\right.\nonumber\\
&+&v_1^2\left(\frac{46}{3}v_1^r v_2^r-\frac{23}{3}{v_1^r}^2-\frac{38}{3} {v_2^r}^2\right)
+v_1\ccdot v_2 \left(\frac{148}{15}v_1^r v_2^r- \frac{79}{15}{v_1^r}^2
+\frac{101}{15}{v_2^r}^2\right)\nonumber\\
&+&\left.v_2^2\left(\frac{79}{30}{v_1^r}^2-\frac{8}{5}v_1^r v_2^r-\frac{67}{10}{v_2^r}^2\right)+v_2^r\left(\frac{332}{15}{v_1^r}^3-\frac{383}{15}{v_1^r}^2v_2^r+\frac{112}{15}v_1^r{v_2^r}^2+\frac{22}{15}{v_2^r}^3\right)\right]r\nonumber\\
&+&a_2^r v_1^2\left[\frac{34}{5}{v_1^r}^2-\frac{229}{15}v_1^r v_2^r+\frac{133}{10}{v_2^r}^2-\frac{121}{30}v_1^2\right]r
+\frac{181}{30}v_1^6-8\left(v_1\ccdot v_2\right)^3+\frac{115}{32}v_2^6+\frac{707}{60}v_1^4v_2^2\nonumber\\
&-&\frac{707}{30}v_1^4 v_1\ccdot v_2+ \frac{43}{2}\left(v_1\ccdot v_2\right)^2 v_1^2+\frac{67}{4}\left(v_1\ccdot v_2\right)^2v_2^2
+\frac{105}{16}v_2^4v_1^2-\frac{105}{8}v_1\ccdot v_2 v_2^4-\frac{43}{2}v_1^2 v_1\ccdot v_2 v_2^2\nonumber\\
&+&v_1^4\left(\frac{326}{15}v_1^r v_2^r-\frac{163}{15}{v_1^r}^2-\frac{157}{10}{v_2^r}^2\right)
+v_2^4\left(\frac{3}{16}{v_2^r}^2-\frac{21}{8}{v_1^r}^2-\frac{3}{8}v_1^r v_2^r\right)
-9\left(v_1\ccdot v_2\right)^2{v_1^r}^2\nonumber\\
&+&v_1^2 v_1\ccdot v_2\left(\frac{128}{5}{v_1^r}^2-\frac{418}{15}v_1^r v_2^r+\frac{314}{15}{v_2^r}^2\right)
+v_1^2 v_2^2\left(\frac{443}{30}v_1^r v_2^r-\frac{64}{5}{v_1^r}^2-\frac{133}{10}{v_2^r}^2\right)\nonumber\\
&+&v_1\ccdot v_2 v_2^2v_1^r\left(9v_1^r+\frac{1}{2}v_2^r\right)+v_1^2\left[\frac{68}{5}{v_1^r}^4-\frac{484}{15}{v_1^r}^3v_2^r+\frac{851}{15}{v_1^r}^2{v_2^r}^2-\frac{724}{15}v_1^r{v_2^r}^3+\frac{266}{15}{v_2^r}^4\right]\nonumber\\
&-&v_1\ccdot v_2 {v_1^r}^3\left(\frac{241}{30}v_1^r+2v_2^r\right)
+v_2^2{v_1^r}^2\left(\frac{5}{4}{v_1^r}^2+v_1^rv_2^r-\frac{1}{2}{v_2^r}^2\right)\nonumber\\
&+&\left.{v_1^r}^4\left(\frac{136}{25}v_1^r v_2^r+\frac{3}{2}{v_2^r}^2-\frac{161}{25}{v_1^r}^2\right)\right\}
+{\cal L}^{a^2}_{G^2_N}\,,
\ee
where terms involving more than one power of accelerations and their
derivatives are:
\be
{\cal L}^{a^2}_{G^2_N}&=&\frac{G_N^2 m_1^2 m_2}{r^2}\left\{\left[\left(\frac{5317}{3600}-\frac{38}{15}\log{\bar{r}}\right)\dot{a}_1 \dot{a}_2
+\left(\frac{4}{15}\log{\bar{r}}-\frac{7}{75}\right)\dot{a}_1^r \dot{a}_2^r-\frac{1}{10}\dot{a}_1 \dot{a}_1 -\frac{14}{45}\dot{a}_1^r\ccdot \dot{a}_1^r\right] r^2\right.\nonumber\\
&+&\left[\left(\frac{68}{15}\log{\bar{r}}-\frac{332}{225}\right)v\ccdot a_1 \dot{a}_2^r-\frac{17}{3}v\ccdot a_1 \dot{a}_1^r
+\left(\frac{24}{5}\log{\bar{r}}-\frac{923}{600}\right)v\ccdot a_2 \dot{a}_1^r\right]r\nonumber\\
&+&a_2\ccdot \dot{a}_1\left[\left(\frac{24}{5}\log{\bar{r}} -\frac{283}{100}\right)v_1^r+\left(\frac{148}{15}\log{\bar{r}} -\frac{3581}{1800}\right)v_2^r\right]r
-r a_2^r \dot{a}_1^r\left(\frac{6}{5}v_1^r+\frac{19}{30}v_2^r\right)\nonumber\\
&+&a_1^2 \left[\left(\frac{38}{15}\log{\bar{r}}-\frac{857}{225}\right) r a_2^r  -\frac{39}{10} r a_1^r
+\left(\frac{44}{3}\log{\bar{r}}-\frac{191}{18}\right)v_1^2+\left(\frac{19}{18}-\frac{22}{3}\log{\bar{r}}\right)v_1\ccdot v_2\right.\nonumber\\
&+&\left.\left(\frac{11}{3}\log{\bar{r}}-\frac{19}{36}\right)v_2^2-\frac{15}{2}{v_1^r}^2+15 v_1^r v_2^r-\frac{23}{6}{v_2^r}^2\right] 
+a_1\ccdot a_2 \left[\left(\frac{28}{3}\log{\bar{r}}+\frac{377}{72}\right) r a_1^r+\right.\nonumber\\
&+&\left.\left(\frac{332}{225}-\frac{68}{15}\log{\bar{r}}\right) r a_2^r+\left(\frac{268}{15}\log{\bar{r}}-\frac{4211}{1800}\right)v_1^2
-\left(\frac{32}{5}\log{\bar{r}}+\frac{124}{25}\right)v_1\ccdot v_2\right.\nonumber\\
&+&\left.\left(\frac{158}{15}\log{\bar{r}}+\frac{253}{3600}\right)v_2^2
+\frac{241}{30}{v_1^r}^2-\frac{107}{30}v_1^r v_2^r+\frac{277}{60}{v_2^r}^2\right] -\frac{2}{3}r {a_1^r}^2 a_2^r
+{a_1^r}^2\left[\frac{1}{6}v_1^2+\frac{2}{3}v_1\ccdot v_2\right.\nonumber\\
&-&\left.\frac{1}{3}v_2^2+\frac{3}{2}{v_1^r}^2-3v_1^r v_2^r+\frac{7}{6}{v_2^r}^2\right]
+a_1^r a_2^r\left[\frac{59}{15}v_1\ccdot v_2-\frac{19}{5}v_1^2-\frac{173}{60}v_2^2+\frac{11}{10}{v_2^r}^2
+\frac{67}{15}v_1^r v_2^r-\frac{56}{15}{v_1^r}^2\right]\nonumber\\
&+&a_1^r\left[\frac{2}{3}v_2\ccdot a_1 v_1^r-\frac{2}{3}v\ccdot a_1v^r+v\ccdot a_2\left(\frac{31}{15}v_1^r+\frac{59}{15}v_2^r\right)-\frac{11}{6}v_2\ccdot a_2 v_2^r\right]\nonumber\\
&+&a_2^r\left[v\ccdot a_1\left(\frac{94}{15}v_1^r-\frac{413}{30}v_2^r\right)-\frac{11}{6}v_2\ccdot a_1 v_1^r\right]
+\left(\frac{22}{3}\log{\bar{r}}-\frac{158}{9}\right)\left(v_1\ccdot a_1\right)^2+\frac{65}{3}v_1\ccdot a_1 v_2\ccdot a_1\nonumber\\
&-&\left.\frac{65}{6}\left(v_2\ccdot a_1\right)^2+\left(\frac{128}{5}\log{\bar{r}}-\frac{267}{200}\right)v\ccdot a_1 v\ccdot a_2
+\left(\frac{44}{3}\log{\bar{r}}-\frac{347}{72}\right)v_2\ccdot a_1 v_1\ccdot a_2 \right\}+\frac{{\cal L}^{pole}}{(d-3)}\,.
\ee
We write separately the divergent terms:
\be
{\cal L}^{pole}&=&-\left(G_N^2 m_1^2 m_2\right)\left\{\left[\frac{2}{15}\dot{a}_1^r \dot{a}_2^r-\frac{19}{15} \dot{a}_1\ccdot \dot{a}_2\right] r^2
+\left[\left(\frac{34}{15}v\ccdot a_1 \dot{a}_2^r+\frac{12}{5}v\ccdot a_2 \dot{a}_1^r\right)
+a_1\ccdot \dot{a}_1\left(\frac{71}{30}v_1^r -\frac{11}{15}v_2^r\right)\right.\right.\nonumber\\
&+&\left.a_2\ccdot \dot{a}_1\left(\frac{12}{5} v_1^r +\frac{74}{15}v_2^r\right)\right]r
+ \left[\frac{71}{60}r a_1^r+\frac{9}{10} r a_2^r+\frac{511}{60} v_1^2-\frac{313}{60} v_1\ccdot v_2+\frac{11}{5} v_2^2\right] a_1^2
+\frac{11}{3}\left(v_1\ccdot a_1\right)^2\nonumber\\
&+&\left.\left[\frac{14}{3}r a_1^r-\frac{34}{15}r a_2^r+\frac{134}{15}v_1^2-\frac{16}{5} v_1\ccdot v_2+\frac{79}{15}v_2^2\right] a_1\ccdot a_2
+\frac{64}{5} v\ccdot a_1 v\ccdot a_2+\frac{22}{3}v_1\ccdot a_2 v_2\ccdot a_1\right\}\,,
\ee
with $v\equiv v_1-v_2$ and $\bar{r}\equiv\sqrt{4 \pi {\rm e}^\gamma}\frac{r}{L}$
($\gamma\simeq 0.577216$ is the Euler-Mascheroni constant).
${\cal L}^{a^2}_{G^2_N}$ is not relevant to determine the 4PN Lagrangian up to 
$G^2_N$ because it gives rise to ${\cal O}(G^3_N)$ terms via the double zero 
trick.
As to its divergent part ${\cal L}^{pole}$, it will have to be canceled by an 
appropriate world-line redefinition, analogously to what happens at 3PN
\cite{Blanchet:2003gy}, thus leaving a finite result in the $d\rightarrow 3$ 
limit.
It is however not possible to determine the exact form of such redefinition at 
this point, as it must be further constrained by the knowledge
of the poles occurring at higher powers of $G_N$.

\section{The 4PN Lagrangian linear in accelerations up to $G^2_N$}
\label{se:linearAcc}
Before we can write the Lagrangian linear in accelerations up to term of order 
$G^2_N$, that is the 4PN extension of the standard formula reported in 
\cite{Blanchet_living}, we have to take into account the manipulations that 
led to such standard formula.
At lower PN orders the double zero trick have been repeatedly used, and its
effect at 4PN order must be computed.

\subsection{2PN double zeros evaluated on 2PN equations of motion}
The following 2PN term quadratic in the accelerations
\be
{\cal L}^{a^2-2PN}=\frac{G_N m_1 m_2 r}{16} \left(15 a_1\cdot a_2 - a_1^r a_2^r\right)
\ee
produces the 2PN and 3PN contributions duly included in the standard 3PN Lagrangian, and the following 4PN term:
\be
\label{doubzPN2G1z2}
\!\!\!\pa{{\cal L}^{a^2-2PN}_{4PN}}_{z^2}\!\!\!\!&=&\!\!\!
\frac{G^2_N m_1^2 m_2}{r^2}\left\{ v_2\ccdot a_1\left[ v_1^2\left(\frac{75}{8}v_1^r-\frac{15}{2} v_2^r\right)-\frac{15}{2}v_1\ccdot v_2v^r-\frac{15}{8}v_2^2 v_1^r+\left(\frac{45}{4}v_2^r-\frac{135}{16}v_1^r\right){v_1^r}^2\right]r \right.\nonumber\\
&+&v_1\ccdot a_1 \left(\frac{15}{2}v_1\ccdot v_2 v^r+\frac{15}{8}v_2^2 v_1^r\right) r+a_1^r\left[\frac{187}{32} v_1^4+ \frac{7}{2} \left(v_1\ccdot v_2\right)^2-7 v_1^2 v_1\ccdot v_2
+v_2^2\left(\frac{1}{8}v_2^r-\frac{11}{4}v_1^r\right)v_1^r\right.\nonumber\\
&+&v_1^2\left(\frac{1}{2} {v_2^r}^2+\frac{81}{8}v_1^r v_2^r-\frac{637}{32} {v_1^r}^2\right)+\left.v_1\ccdot v_2\left(10 {v_1^r}^2+v_1^rv_2^r-\frac{1}{2}{v_2^r}^2\right)+\left(\frac{21}{16}v_1^r-\frac{3}{4}v_2^r\right){v_1^r}^2 v_2^r\right]\nonumber\\
&+&a_2^r v_1^2\left(\frac{45}{8}{v_1^r}^2-\frac{15}{8}v_1^2\right)+\frac{75}{32}v_1^6-\frac{135}{32}v_1^4 v_1\ccdot v_2+\frac{15}{8}v_1^4 v_2^2
+v_1^4\left(\frac{315}{16}v_1^r v_2^r-\frac{555}{32}{v_1^r}^2-\frac{15}{4}{v_2^r}^2\right)\nonumber\\
&+&v_1^2 v_1\ccdot v_2\left(\frac{585}{32}v_1^r-\frac{45}{4}v_2^r\right)v_1^r-\frac{45}{8}v_1^2 v_2^2 {v_1^r}^2+v_1^2\left(\frac{453}{32}{v_1^r}^2-\frac{315}{8}v_1^r v_2^r+\frac{45}{2}{v_2^r}^2\right){v_1^r}^2\nonumber\\
&+&\left.\frac{87}{32}v_1\ccdot v_2{v_1^r}^4+\frac{261}{80}v^r {v_1^r}^5\right\}+{\cal O}(G^3_N)\,.
\ee
\subsection{3PN double zeros evaluated on 1PN equations of motion}
The following 3PN terms quadratic in accelerations
\be
{\cal L}^{a^2-3PN}&=&\left(G_N m_1 m_2\right) r \left\{\left(\frac{1}{96} \dot{a}_1^r \dot{a}_2^r -\frac{23}{288} \dot{a}_1\ccdot \dot{a}_2\right)r^2+
\left(\frac{15}{32}a_2\ccdot \dot{a}_1 -\frac{1}{32} a_2^r \dot{a}_1^r\right)\left(v_1^r+v_2^r\right) r+\frac{7}{16} r v\ccdot a_2 \dot{a}_1^r \right.\nonumber\\
&-&\frac{r}{32}{a_1^r}^2 a_2^r+a_1\ccdot a_2\left(\frac{29}{32} r a_1^r -\frac{5}{16} v_1\ccdot v_2+ \frac{55}{32}v_1^2+ \frac{29}{32} {v_1^r}^2+ \frac{7}{16} v_1^r v_2^r\right)
\nonumber\\
&+&a_1^r a_2^r\left(\frac{3}{16} v_1\ccdot v_2-\frac{9}{32} v_1^2+ \frac{1}{16}v_1^r v_2^r+ \frac{3}{32} {v_1^r}^2\right)
+a_1^r v\ccdot a_2\left(\frac{41}{32} v_1^r+\frac{13}{32}v_2^r\right)-\frac{1}{8} a_1^r v_2\ccdot a_2 v_2^r\nonumber\\
&+&\left.\frac{47}{16} v_1\ccdot a_1 v_1\ccdot a_2-\frac{47}{32} v_1\ccdot a_1 v_2\ccdot a_2-\frac{17}{32} v_2\ccdot a_1 v_1\ccdot a_2\right\}
\ee
generate (as well as several 3PN terms) the following contribution:
\be
\label{doubzPN3G1z2}
\!\pa{{\cal L}^{a^2-3PN}_{4PN}}_{z^2}&=&\!
\frac{G^2_N m_1^2 m_2}{r^2}\left\{v_1\ccdot a_1 \left[v_1\ccdot v_2 \left(\frac{245}{24}v_1^r-\frac{271}{24}v_2^r\right)+v_2^2 \left(\frac{407}{24}v_2^r-\frac{71}{8}v_1^r\right)\right] r\right. \nonumber\\
&+&v_2\ccdot a_1\left[ v_1^2\left(\frac{355}{48}v_1^r-\frac{35}{24} v_2^r\right)+v_1\ccdot v_2\left(\frac{91}{24}v_2^r-\frac{139}{12}v_1^r\right)
+v_2^2\left(\frac{85}{8}v_1^r-\frac{407}{24}v_1^r\right) +\frac{61}{24}{v_1^r}^3\right.\nonumber\\
&-&\left.\frac{183}{16}{v_1^r}^2v_2^r-\frac{181}{48}v_1^r {v_2^r}^2+\frac{23}{12}{v_2^r}^3\right]r
+a_1^r\left[\frac{689}{192} v_1^4+ \frac{17}{6} v_2^4-\frac{21}{8}v_1^2 v_1\ccdot v_2+\frac{16}{3}v_1^2 v_2^2\right.\nonumber\\
&-&\frac{75}{8}v_1\ccdot v_2 v_2^2+v_1^2\left(\frac{967}{48}{v_1^r}^2-\frac{55}{6}v_1^rv_2^r-\frac{337}{96}{v_2^r}^2\right)
+v_1\ccdot v_2\left(\frac{67}{24} {v_2^r}^2-\frac{31}{6}v_1^rv_2^r-\frac{253}{12}{v_1^r}^2\right)\nonumber\\
&+&\left.v_2^2\left(\frac{35}{6}v_1^r v_2^r-\frac{3}{2} {v_1^r}^2-\frac{2}{3} {v_2^r}^2\right)
+\left(\frac{439}{48}{v_1^r}^3+\frac{193}{48}{v_1^r}^2v_2^r-\frac{33}{16}v_1^r {v_2^r}^2-\frac{1}{4}{v_2^r}^3\right)v_2^r\right]\nonumber\\
&+&a_2^r v_1^2\left(\frac{23}{8}{v_2^r}^2-\frac{205}{48}v_1^r v_2^r-\frac{169}{32}{v_1^r}^2-\frac{215}{96}v_1^2\right)
+\frac{457}{192}v_1^6-\frac{887}{192}v_1^4 v_1\ccdot v_2+\frac{215}{96}v_1^4 v_2^2\nonumber\\
&+&v_1^4\left(\frac{233}{96}{v_1^r}^2-\frac{127}{96}v_1^r v_2^r-\frac{635}{96}{v_2^r}^2\right)
+v_1^2 v_1\ccdot v_2\left(\frac{481}{96}{v_2^r}^2+\frac{151}{24}v_1^r v_2^r-\frac{339}{32}{v_1^r}^2\right)\nonumber\\
&+&v_1^2 v_2^2\left(\frac{169}{32}{v_1^r}^2+\frac{205}{48}v_1^r v_2^r-\frac{23}{8}{v_2^r}^2\right)
-\frac{91}{6}v_1\ccdot v_2{v_1^r}^4\nonumber\\
&+&\left.v_1^2\left(\frac{67}{12}{v_1^r}^4+\frac{737}{24}{v_1^r}^3 v_2^r-\frac{151}{12}{v_1^r}^2{v_2^r}^2-\frac{99}{8}v_1^r {v_2^r}^2+\frac{23}{6}{v_2^r}^4\right)
-\frac{91}{5}v^r {v_1^r}^5\right\}+{\cal O}(G^3_N)\,.
\ee
\subsection{4PN double zeroes from $O(G)$ graphs}
Finally, acting with the double zero trick on ${\cal L}^{a^2}_{G_N}$ 
reported in eq.(\ref{doubzG1}), gives 
\be
\label{doubzPN4G1z2}
\!\pa{{\cal L}^{a^2}_{G_N}}_{z^2}&=&\frac{G_N^2 m_1^2 m_2}{r^2}\left\{v_1\ccdot a_1\left[v_1\ccdot v_2\left(\frac{65}{48}v_1^r+\frac{137}{48}v_2^r\right)
-v_2^2 \left(\frac{81}{32}v_1^r+\frac{665}{96}v_2^r\right)\right]r\right.\nonumber\\
&+&v_2\ccdot a_1\left[v_1^2\left(\frac{95}{96}v_2^r-\frac{65}{96} v_1^r\right)+ v_1\ccdot v_2\left(\frac{103}{48}v_1^r+\frac{25}{48}v_2^r\right)
+v_2^2 \left(\frac{113}{32}v_1^r+\frac{665}{96}v_2^r\right)\right.\nonumber\\
&+&\left.\frac{557}{96}{v_1^r}^3+\frac{39}{32}{v_1^r}^2v_2^r-\frac{121}{96}v_1^r{v_2^r}^2-\frac{37}{32}{v_2^r}^3\right]r\nonumber\\
&+&a_1^r\left[\frac{185}{384}v_1^4+\frac{23}{16}\left(v_1\ccdot v_2\right)^2+\frac{89}{192}v_2^4-\frac{13}{16}v_1^2 v_1\ccdot v_2
-\frac{11}{96} v_1^2v_2^2+\frac{27}{16}v_1\ccdot v_2 v_2^2\right.\nonumber\\
&+&v_1^2\left(\frac{499}{96} v_1^r v_2^r-\frac{673}{192}{v_1^r}^2
-\frac{49}{192} {v_2^r}^2\right)+v_1\ccdot v_2\left(\frac{52}{3}{v_1^r}^2-\frac{73}{12}v_1^r v_2^r-\frac{19}{24} {v_2^r}^2\right)\nonumber\\
&+&\left.v_2^2\left(\frac{9}{8}{v_1^r}^2-\frac{71}{24}v_1^r v_2^r+\frac{5}{12}{v_2^r}^2\right)
+v_2^r\left(\frac{805}{96}{v_1^r}^2v_2^r+\frac{1}{6}v_1^r{v_2^r}^2-\frac{375}{16}{v_1^r}^3-\frac{7}{192} {v_2^r}^3\right)\right]r\nonumber\\
&+&a_2^r v_1^2\left[\frac{419}{384}v_1^2+\frac{147}{64}{v_1^r}^2-\frac{121}{96} v_1^r v_2^r-\frac{111}{64} {v_2^r}^2\right]r
-\frac{53}{384}v_1^6+\frac{59}{48}v_1^4 v_1\ccdot v_2-\frac{419}{384}v_1^4v_2^2\nonumber\\
&+&v_1^4\left(\frac{149}{96}{v_2^r}^2+\frac{205}{96}v_1^r v_2^r-\frac{19}{48}{v_1^r}^2\right)
+v_1^2v _1\ccdot v_2\left(\frac{95}{32}{v_1^r}^2-\frac{281}{48}v_1^r v_2^r-\frac{53}{48}{v_2^r}^2\right)\nonumber\\
&+&v_1^2v_2^2\left(\frac{111}{64}{v_2^r}^2+\frac{121}{96}v_1^r v_2^r-\frac{147}{64}{v_1^r}^2\right)+\frac{2615}{192} v_1\ccdot v_2 {v_1^r}^4\nonumber\\
&+&\left.v_1^2\left(\frac{281}{24}{v_1^r}^2{v_2^r}^2-\frac{2443}{192}{v_1^r}^4-\frac{121}{12}{v_1^r}^3v_2^r-\frac{5}{24}v_1^r{v_2^r}^3-\frac{37}{16}{v_2^r}^4\right)+\frac{523}{32} v^r{v_1^r}^5\right\}
+{\cal O}(G^3_N)\,.\nonumber\\
\ee

\subsection{Result of the double zero procedure}
Summing up all the contributions, we have:
\be
{\cal L}^{4PN}=\frac{7}{256} v_1^{10}+\hat{{\cal L}}^{4PN}_{G_N}+\hat{{\cal L}}^{4PN}_{G^2_N}+{\cal O}(G^3_N)\,,
\ee
where $\hat{{\cal L}}^{4PN}_{G_N}\equiv{\cal L}^{4PN}_{G_N}-{\cal L}^{a^2}_{G_N}$ can 
be read directly in eq.(\ref{G1tot}), while $\hat{{\cal L}}^{4PN}_{G^2_N}$ 
receives contributions from the 
eqs.(\ref{doubzPN2G1z2},\ref{doubzPN3G1z2},\ref{doubzPN4G1z2}), and reads:
\be\label{laG2tot}
\hat{{\cal L}}^{4PN}_{G^2_N}&=&\frac{G_N^2 m_1^2 m_2}{r^2}\left\{v_1\ccdot a_1\left[v_2^2 \left(\frac{3733}{160}v_2^r-\frac{3679}{480}v_2^r\right)
+v_1\ccdot v_2\left(\frac{3679}{240}v_1^r-\frac{8849}{240}v_2^r\right)\right]r\right.\nonumber\\
&+&v_2\ccdot a_1\left[v_1^2\left(\frac{9229}{480}v_1^r -\frac{8849}{480}v_2^r\right)+ v_1\ccdot v_2\left(\frac{6499}{240}v_2^r-\frac{4529}{240} v_1^r\right)\right.\nonumber\\
&+&\left.v_2^2 \left(\frac{2293}{160}v_1^r-\frac{3733}{160}v_2^r\right)
-\frac{3341}{480}{v_1^r}^3+\frac{10223}{480}{v_1^r}^2v_2^r-\frac{3781}{160}v_1^r{v_2^r}^2+\frac{4621}{480}{v_2^r}^3\right]r\nonumber\\
&+&a_1^r\left[\frac{6943}{384}v_1^4+\frac{1331}{80}\left(v_1\ccdot v_2\right)^2+\frac{2931}{320}v_2^4
+\frac{7193}{480}v_1^2v_2^2-\frac{7193}{240}v_1^2v_1\ccdot v_2-\frac{5593}{240}v_1\ccdot v_2v_2^2\right.\nonumber\\
&+&v_1^2\left(\frac{2063}{96}v_1^r v_2^r-\frac{2099}{192}{v_1^r}^2-\frac{3059}{192} {v_2^r}^2\right)
+v_1\ccdot v_2 \left( \frac{59}{60}{v_1^r}^2-\frac{23}{60}v_1^r v_2^r+\frac{247}{30}{v_2^r}^2\right)\nonumber\\
&+&\left.v_2^2\left(\frac{7}{5}v_1^r v_2^r-\frac{59}{120}{v_1^r}^2-\frac{139}{20}{v_2^r}^2\right)
+v_2^r\left(\frac{2197}{240}{v_1^r}^3-\frac{6661}{480}{v_1^r}^2v_2^r+\frac{1337}{240}v_1^r{v_2^r}^2+\frac{1133}{960}{v_2^r}^3\right)\right]r\nonumber\\
&+&a_2^r v_1^2\left[\frac{3021}{320}{v_1^r}^2-\frac{9983}{480}v_1^r v_2^r+\frac{4621}{320}{v_2^r}^2-\frac{13549}{1920}v_1^2\right]r
+\frac{20389}{1920}v_1^6-8\left(v_1\ccdot v_2\right)^3+\frac{115}{32}v_2^6\nonumber\\
&+&\frac{28429}{1920}v_1^4v_2^2-\frac{29929}{960}v_1^4 v_1\ccdot v_2+\frac{43}{2}\left(v_1\ccdot v_2\right)^2  v_1^2+\frac{67}{4}\left(v_1\ccdot v_2\right)^2v_2^2
+\frac{105}{16}v_2^4v_1^2-\frac{105}{8}v_1\ccdot v_2 v_2^4\nonumber\\
&-&\frac{43}{2}v_1^2 v_1\ccdot v_2 v_2^2+v_1^4\left(\frac{1267}{30}v_1^r v_2^r-\frac{6283}{240}{v_1^r}^2-\frac{1961}{80}{v_2^r}^2\right)
+v_2^4\left(\frac{3}{16}{v_2^r}^2-\frac{21}{8}{v_1^r}^2-\frac{3}{8}v_1^r v_2^r\right)\nonumber\\
&-&9\left(v_1\ccdot v_2\right)^2{v_1^r}^2+v_1^2 v_1\ccdot v_2\left(\frac{5501}{160}{v_1^r}^2-\frac{9283}{240}v_1^r v_2^r+\frac{11923}{480}{v_2^r}^2\right)
+v_1\ccdot v_2 v_2^2v_1^r\left(9v_1^r+\frac{1}{2}v_2^r\right)\nonumber\\
&+&v_1^2 v_2^2\left(\frac{9743}{480}v_1^r v_2^r-\frac{4941}{320}{v_1^r}^2-\frac{4621}{320}{v_2^r}^2\right)
+v_2^2{v_1^r}^2\left(\frac{5}{4}{v_1^r}^2+v_1^rv_2^r-\frac{1}{2}{v_2^r}^2\right)\nonumber\\
&+&v_1^2\left[\frac{6597}{320}{v_1^r}^4-\frac{3061}{60}{v_1^r}^3v_2^r+\frac{9403}{120}{v_1^r}^2{v_2^r}^2-\frac{1217}{20}v_1^r{v_2^r}^3+\frac{4621}{240}{v_2^r}^4\right]
-v_1\ccdot v_2 {v_1^r}^3\left(\frac{6587}{960}v_1^r+2v_2^r\right)\nonumber\\
&+&\left.{v_1^r}^4\left(\frac{3227}{800}v_1^r v_2^r+\frac{3}{2}{v_2^r}^2-\frac{4027}{800}{v_1^r}^2\right)\right\}\,.
\ee
The above formula, along with the term explicitly written in equation (\ref{G1tot}), is the main result of this paper.

\section{Energy}
\label{se:energy}
The following expression for the Energy is derived from the Lagrangian via a 
generalized Legendre transform :
\be
E^{4PN}&=&\frac{63}{256}m_1 v_1^{10}+\frac{G_N m_1 m_2}{r}\left\{\frac{525}{128}v_1^8+\frac{1}{32}\left(v_1\ccdot v_2\right)^4-\frac{1291}{128}v_1^6v_1\ccdot v_2
+\frac{731}{128}v_1^6v_2^2+\frac{3}{32}v_1^2\left(v_1\ccdot v_2\right)^3\right.\nonumber\\
&+&\frac{373}{64}v_1^4\left(v_1\ccdot v_2\right)^2+\frac{765}{256}v_1^4v_2^4-\frac{1511}{128}v_1^4v_1\ccdot v_2v_2^2
+3 v_1^2 \left(v_1\ccdot v_2\right)^2v_2^2+\left(v_1\ccdot v_2\right)^3\left(\frac{3}{32}v_1^r+\frac{1}{8}v_2^r\right)v_1^r\nonumber\\
&-&v_1^6\left(\frac{181}{128}v_1^r+\frac{295}{128}v_2^r\right)v_2^r+v_1^4 v_1\ccdot v_2\left(\frac{501}{128}{v_1^r}^2+\frac{35}{32}v_1^r v_2^r+\frac{451}{128}{v_2^r}^2\right)\nonumber\\
&-&v_1^4 v_2^2\left(\frac{501}{128}{v_1^r}^2+\frac{185}{128}v_1^r v_2^r+\frac{231}{64}{v_2^r}^2\right)
+v_1^2 \left(v_1\ccdot v_2\right)^2\left(\frac{23}{32}v_1^r v_2^r-\frac{51}{32}{v_1^r}^2-\frac{21}{16}{v_2^r}^2\right)\nonumber\\
&+&v_1^2 v_1\ccdot v_2 v_2^2\left(\frac{333}{64}v_1^r+\frac{3}{8} v_2^r\right)v_1^r+v_1^4\left(\frac{267}{128}{v_1^r}^3+\frac{159}{128}{v_1^r}^2v_2^r+\frac{93}{128}v_1^r{v_2^r}^2
+\frac{183}{128}{v_2^r}^3\right)v_2^r\nonumber\\
&+&\left(v_1\ccdot v_2\right)^2\left(\frac{31}{64}{v_1^r}^2-\frac{25}{32}v_1^rv_2^r-\frac{15}{32}{v_2^r}^2\right){v_1^r}^2
+v_1^2 v_2^2\left(\frac{411}{128}{v_1^r}^2+\frac{99}{64}v_1^r v_2^r+\frac{3}{4}{v_2^r}^2\right){v_1^r}^2\nonumber\\
&-&v_1^2 v_1\ccdot v_2\left(\frac{411}{128}{v_1^r}^4+\frac{25}{16}{v_1^r}^3 v_2^r+\frac{45}{64}{v_1^r}^2{v_2^r}^2+\frac{3}{8}v_1^r{v_2^r}^3+\frac{245}{128}{v_2^r}^4\right)
+v_1\ccdot v_2\left(\frac{125}{128}{v_1^r}^3+\frac{25}{32}{v_1^r}^2v_2^r\right.\nonumber\\
&+&\left.\frac{85}{128}v_1^r{v_2^r}^2+\frac{5}{16}{v_2^r}^3\right){v_1^r}^3-v_1^2\left(\frac{165}{128}{v_1^r}^5+\frac{135}{128}{v_1^r}^4v_2^r
+\frac{55}{64}{v_1^r}^3 {v_2^r}^2+\frac{45}{64}{v_1^r}^2{v_2^r}^3+\frac{75}{128}v_1^r{v_2^r}^4\right.\nonumber\\
&+&\left.\left.\frac{125}{128}{v_2^r}^5\right)v_2^r+\frac{35}{128}\left({v_1^r}^3+{v_1^r}^2v_2^r
+v_1^r{v_2^r}^2+\frac{1}{2}{v_2^r}^3\right){v_1^r}^4v_2^r\right\}\nonumber\\
&+&\frac{G^2_N m_1^2 m_2}{r^2}\left\{\frac{45}{2}v_1^6-65\left(v_1\ccdot v_2\right)^3+\frac{575}{32}v_2^6-104v_1^4v_1\ccdot v_2 +\frac{201}{4}v_1^4v_2^2
+\frac{293}{2}v_1^2\left(v_1\ccdot v_2\right)^2\right.\nonumber\\
&+&\frac{469}{4}\left(v_1\ccdot v_2\right)^2v_2^2+\frac{691}{16}v_1^2 v_2^4-76v_1\ccdot v_2v_2^4-\frac{305}{2}v_1^2 v_1\ccdot v_2 v_2^2
+v_1^4\left(\frac{5451}{64}v_1^rv_2^r-\frac{2171}{64}{v_1^r}^2-\frac{175}{4}{v_2^r}^2\right)\nonumber\\
&+&\left(v_1\ccdot v_2\right)^2\left(\frac{2025}{16}v_1^rv_2^r-\frac{1461}{16}{v_1^r}^2-77{v_2^r}^2\right)
+v_2^4\left(\frac{295}{64}{v_1^r}^2-\frac{1543}{64}v_1^rv_2^r+\frac{15}{16}{v_2^r}^2\right)\nonumber\\
&+&v_1^2v_1\ccdot v_2\left(\frac{1805}{16}{v_1^r}^2-\frac{3331}{16}v_1^rv_2^r+\frac{483}{4}{v_2^r}^2\right)
+v_1^2v_2^2\left(\frac{2021}{32}v_1^rv_2^r-\frac{949}{32}{v_1^r}^2-\frac{371}{8}{v_2^r}^2\right)\nonumber\\
&+&v_1\ccdot v_2v_2^2\left(\frac{817}{16}{v_1^r}^2-\frac{779}{16}v_1^rv_2^r+\frac{371}{8}{v_2^r}^2\right)
+v_1^2\left(\frac{769}{32}{v_1^r}^4+\frac{4683}{32}{v_1^r}^2{v_2^r}^2-\frac{3735}{32}{v_1^r}^3v_2^r\right.\nonumber\\
&-&\left.\frac{10615}{96}{v_2^r}^3v_1^r+\frac{125}{4}{v_2^r}^4\right)+v_1\ccdot v_2\left(\frac{2659}{16}{v_1^r}^3v_2^r-\frac{2969}{16}{v_1^r}^2{v_2^r}^2+\frac{1741}{16}v_1^r{v_2^r}^3-\frac{125}{4}{v_2^r}^4-\frac{811}{16}{v_1^r}^4\right)\nonumber\\
&+&v_2^2\left(\frac{363}{32}{v_1^r}^3-\frac{833}{32}{v_1^r}^2v_2^r-\frac{615}{32}v_1^r{v_2^r}^2+\frac{2639}{96}{v_2^r}^3\right)v_1^r
+\left(\frac{4739}{64}{v_1^r}^4v_2^r-\frac{3819}{320}{v_1^r}^5\right.\nonumber\\
&-&\left.\left.\frac{4223}{32}{v_1^r}^3{v_2^r}^2+\frac{3775}{32}{v_1^r}^2{v_2^r}^3-\frac{1699}{64}v_1^r{v_2^r}^4-\frac{1941}{320}{v_2^r}^5\right)v_1^r\right\}+{\cal O}(G_N^3)\,.
\ee
In the above expression (which is conserved up to order $G^2$ if gravitational 
radiation is neglected) accelerations have been systematically
substituted via the equations of motion, truncated at the appropriate
PN order.

\section{Lorentz invariance and center-of-mass position}
\label{se:lorentz}
\subsection{Boosts}
As discussed in \cite{deAndrade:2000gf}, the post-Newtonian dynamics must 
inherit the symmetries of the fundamental theory, thus has to show invariance 
under Lorentz boosts, which in this framework are implemented by the following 
transformation
\be
\label{boost}
\delta \vec{x}_{1,2}=-\vec{V} t+\left(V\cdot x_{1,2}\right)\vec{v}_{1,2}+{\cal O}(V^2)\,,
\ee
where $V^i$ is the boost velocity.

Requiring that the equations of motion are invariant under boosts impose
that the Lagrangian variation under the transformation given by eq.~(\ref{boost})
consists only of a time derivative plus possibly double zero terms:
\be
\label{varL}
\delta {\cal L}^{4PN}=V\cdot \dot{Z}^{4PN}+{\rm double\ zeroes}\,.
\ee
As the boost transformation does not depend on $G_N$ (nor on $d$), Lorentz 
invariance can be checked to some extent order by order in Newton's constant;
we find indeed that the above eq.~(\ref{varL}) is compatible with 
$\vec{Z}^{4PN}=\frac{5}{128}v_1^8 \vec{x}_1+\vec{Z}^{4PN}_{G_N}+\vec{Z}^{4PN}_{G^2_N}+{\cal O}(G^3_N)$, where
\be
\vec{Z}^{4PN}_{G_N}&=&\frac{G m_1 m_2}{r}\left\{\left[\frac{11}{4}v_1^4 v_2^r+\left(v_1\ccdot v_2\right)^2\left(\frac{1}{4}v_2^r-\frac{3}{16}v_1^r\right)
+v_2^4\left(\frac{49}{64}v_1^r+\frac{123}{64}v_2^r\right)-3v_1^2v_1\ccdot v_2 v_2^r\right.\right.\nonumber\\
&-&\frac{123}{32}v_1^2v_2^2 v_1^r+v_1\ccdot v_2\left(\frac{19}{32}v_1^r{v_2^r}^2-\frac{5}{32}{v_1^r}^2v_2^r+\frac{5}{12}{v_2^r}^3
-\frac{1}{16}{v_1^r}^3\right)+v_2^2\left(\frac{23}{32}v_2^r-\frac{77}{96}v_1^r\right){v_1^r}^2\nonumber\\
&-&\left.\frac{1}{16}v_2^2v_1\ccdot v_2 v_1^r+\left(\frac{13}{64}{v_1^r}^4+\frac{11}{64}{v_1^r}^3v_2^r+\frac{5}{32}{v_1^r}^2{v_2^r}^2
+\frac{1}{8}v_1^r{v_2^r}^3+\frac{1}{16}{v_2^r}^4\right)v_1^r\right]\vec{v_1}\nonumber\\
&+&\left[v_2\ccdot a_1\left(\frac{1}{4}v_1\ccdot v_2-v_1^2-\frac{15}{8}v_2^2+\frac{3}{8}v_1^r v_2^r+\frac{5}{12}{v_2^r}^2\right)v_2^r
+v_1\ccdot a_1\left(\frac{11}{4}v_1^2-2v_1\ccdot v_2+\frac{15}{8}v_2^2\right)v_2^r\right.\nonumber\\
&+&a_1^r\left(\frac{3}{8}v_1\ccdot v_2{v_2^r}^2-v_2^2\left\{\frac{5}{8}{v_1^r}^2+\frac{1}{2}v_1^r v_2^r+\frac{5}{16}{v_2^r}^2\right\}
+\frac{1}{8}v_1^r{v_2^r}^3+\frac{1}{16}{v_2^r}^4\right)+\frac{11}{16}v_1^6+\frac{1}{16}\left(v_1\ccdot v_2\right)^2\nonumber\\
&-&\frac{15}{16}v_1^4v_2^2-\frac{13}{16}v_1^4 v_1\ccdot v_2+v_1^2v_2^4-\frac{1}{32}v_1^2 v_1\ccdot v_2 v_2^2
-v_1^4\left(\frac{7}{16}v_2^r+\frac{11}{16}v_1^r\right)v_2^r+v_1^2 v_1\ccdot v_2\left(\frac{11}{8}v_2^r+v_1^r\right)v_2^r\nonumber\\
&-&v_1^2v_2^2\left(\frac{5}{8}v_1^r+\frac{23}{32}v_2^r\right)v_1^r
+\left(v_1\ccdot v_2\right)^2\left(\frac{11}{64}{v_2^r}^2-\frac{11}{64}{v_1^r}^2+\frac{5}{16}v_1^rv_2^r\right)
-v_1\ccdot v_2 v_2^2{v_1^r}^2+\frac{11}{16}v_2^4{v_1^r}^2\nonumber\\
&+&\left.\left.v_1^2\left(\frac{1}{8}v_1^r+\frac{5}{8}v_2^r\right){v_2^r}^3
+v_1\ccdot v_2\left(\frac{19}{16}{v_1^r}^4+{v_1^r}^3v_2^r-\frac{15}{32}{v_1^r}^2{v_2^r}^2-v_1^r{v_2^r}^3-\frac{19}{16}{v_2^r}^4\right)
-\frac{5}{32}{v_1^r}^3{v_2^r}^3\right]\vec{x}_1\right\}\nonumber\,,\\
\ee
and
\be
\vec{Z}^{4PN}_{G^2_N}&=&\frac{G^2 m_1 m_2}{r^2}\left\{m_2\left[\frac{235}{24}v_1\ccdot v_2 v_2^r+\frac{235}{48}v_2^2 v_1^r+\frac{29}{24}v_1^r{v_2^r}^2\right]\vec{v_1}\right.\nonumber\\
&+&m_1\left[v_1^2\left(\frac{3679}{480}v_1^r-\frac{8849}{480}v_2^r\right)
+\frac{235}{12}v_1\ccdot v_2 v_1^r-\frac{3341}{480}{v_1^r}^3+\frac{10223}{480}{v_1^r}^2v_2^r-\frac{9983}{480}v_1^r{v_2^r}^2+\frac{4621}{480}{v_2^r}^3\right]\!\!\vec{v_1}\nonumber\\
&+&m_1\left[\frac{235}{24}v_2\ccdot a_1 v_2^r+a _1^r\left(\frac{185}{8}v_1\ccdot v_2-\frac{185}{16}v_1^2-\frac{20}{3}v_2^2
+\frac{17}{6}{v_2^r}^2\right)-\frac{149}{24}v_1^4+\frac{235}{16}v_1^2 v_2^2\right.\nonumber\\
&+&\left.v_1^2\left(\frac{161}{24}{v_1^r}^2-\frac{223}{24}v_1^rv_2^r+\frac{97}{24}{v_2^r}^2\right)-\frac{147}{8}v_1\ccdot v_2 {v_1^r}^2
+\left(\frac{10841}{2880}v_1^r-\frac{29}{6}v_2^r\right){v_1^r}^3\right]\vec{x_1}\nonumber\\
&+&m_2\left[-\frac{235}{24}v_2\ccdot a_1 v_2^r-a_1^r\left(\frac{235}{48}v_2^2+\frac{29}{24}{v_2^r}^2\right)
-\frac{45}{16}v_1^4-\frac{463}{24}\left(v_1\ccdot v_2\right)^2-\frac{25}{16}v_2^4+\frac{19}{2}v_1^2 v_1\ccdot v_2\right.\nonumber\\
&-&\frac{463}{48}v_1^2 v_2^2+\frac{733}{48}v_1\ccdot v_2 v_2^2+v_1^2\left(\frac{1}{2}v_1^r v_2^r-\frac{1}{4}{v_1^r}^2+\frac{7}{24}{v_2^r}^2\right)
+v_2^2\left(\frac{23}{4}{v_1^r}^2-\frac{45}{8}v_1^r v_2^r+\frac{7}{2}{v_2^r}^2\right)\nonumber\\
&+&\left.\left.v_1\ccdot v_2\left(\frac{97}{6}v_1^r-\frac{11}{3}v_2^r\right)v_2^r
+\left(\frac{35}{6}{v_1^r}^2+\frac{2}{9}v_1^r v_2^r-\frac{4187}{960}{v_2^r}^2\right){v_2^r}^2\right]\vec{x_1}\right\}\,,
\ee
where the sum over particle exchange $1\leftrightarrow 2$ is understood as 
usual.

More exactly, we find that eq.(\ref{varL}) holds up to order $G_N$ if we 
assume that the following terms
\be
\label{asquareG1}
&&\frac{G_N m_1 m_2}{r}\left[a_1\ccdot a_2\left(\frac{1}{4}v_1\ccdot v_2-v_2^2-\frac{15}{8}v_1^2+\frac{3}{8}v_1^r v_2^r+\frac{5}{12}{v_1^r}^2\right)v_1^r
+\frac{15}{4}v_1\ccdot a_1 v_2\ccdot a_2 v_1^r+\right.\nonumber\\
&&+\frac{1}{4}v_2\ccdot a_1 v_1\ccdot a_2 v_1^r+v_1\ccdot a_1 a_2^r\left(\frac{11}{4}v_1^2-2 v_1\ccdot v_2+\frac{15}{8}v_2^2-\frac{5}{8}{v_1^r}^2-v_1^r v_2^r-\frac{5}{4}v_2^r\right)\nonumber\\
&&+v_2\ccdot a_1 a_2^r\left(\frac{1}{4}v_1\ccdot v_2-v_1^2-\frac{15}{8}v_2^2+\frac{3}{8}{v_1^r}^2+\frac{3}{4}v_1^r v_2^r+\frac{5}{4}v_2^r\right)\nonumber\\
&&+\left.a_1^r a_2^r\left(\frac{3}{4}v_1\ccdot v_2-\frac{5}{8}v_1^2-\frac{1}{2}v_2^2+\frac{1}{4}{v_1^r}^2+\frac{3}{8}v_1^r v_2^r\right)v_1^r-v_1\ccdot a_1 v_1\ccdot a_2 \left(\frac{15}{4}v_1^r+2v_2^r\right)\right]\vec{r}
\ee
are part of the double zero structure displayed at the right hand side of the same equation.
Although we cannot be sure that this is indeed the case without knowing also 
the $G^3_N$ part of the Lagrangian, it is non-trivial consistency check that 
only terms quadratic in the accelerations appear above and that the $G^2_N$
terms are also consistent with the presence of a double zero
structure whose terms quadratic in the accelerations are the ones
given in (\ref{asquareG1}).
 
In its turn, Lorentz invariance at $G^2_N$ order requires
the following ${\cal O}(G^2_N)$ terms quadratic in acceleration
\be
&&\frac{G_N^2m_1^2 m_2}{r^2}\left\{\left[a_1^r a_2^r\left(\frac{29}{12}v_1^r-\frac{17}{3}v_2^r\right)+\frac{235}{24}\left(a_2^r v\ccdot a_1 +v^r a_1\ccdot a_2\right)-\frac{40}{3}v\ccdot a_2 a_1^r\right]\vec{r}\right\}\nonumber\\
\ee
are part of a double zero structure. While an actual check will be performed
only after the determination of the 4PN $G^3_N$ and $G^4_N$
dynamics, it is however a non-trivial that only terms quadratic in 
accelerations are left out of $Z^{4PN}_{G_N^2}$.
Note that in deriving these results we had to take into account the 1PN 
corrections to the double zero terms which are already present at 3PN.

\subsection{Center of mass position}
Still following \cite{deAndrade:2000gf}, we can use the result of the previous subsection, and the fact of having a Lagrangian linear in acceleration, to determine the center-of-mass position at order
$G^2_N$. \be
\vec{G}^{4PN}&=&\frac{35}{128}m_1 v_1^8\vec{x}_1+\frac{G_N m_1m_2}{r}\left\{\left[\frac{515}{128}v_1^6+\frac{1}{16}\left(v_1\ccdot v_2\right)^3-\frac{433}{64}v_1^4 v_1\ccdot v_2+\frac{381}{128}v_1^4v_2^2
+\frac{251}{128}v_1^2v_2^4\right.\right.\nonumber\\
&-&\frac{161}{32}v_1^2v_1\ccdot v_2 v_2^2-\frac{75}{128}v_2^6+\frac{97}{32}v_1^2 \left(v_1\ccdot v_2\right)^2+\frac{67}{32}\left(v_1\ccdot v_2\right)^2 v_2^2-\frac{123}{64}v_1\ccdot v_2 v_2^4\nonumber\\
&-&v_1^4\left(\frac{53}{128}{v_1^r}^2+\frac{31}{64}v_1^r v_2^r+\frac{225}{128}{v_2^r}^2\right)+\left(v_1\ccdot v_2\right)^2\left(\frac{5}{32}{v_1^r}^2+\frac{5}{16}v_1^r v_2^r-\frac{29}{32}{v_2^r}^2\right)\nonumber\\
&+&v_2^4\left(\frac{53}{128}{v_2^r}^2-\frac{53}{64}v_1^r v_2^r-\frac{31}{128}{v_2^r}^2\right)
-v_1^2 v_2^2\left(\frac{23}{64}{v_1^r}^2+\frac{23}{32}v_1^r v_2^r+\frac{157}{64}{v_2^r}^2\right)\nonumber\\
&+&v_1^2 v_1\ccdot v_2\left(\frac{13}{32}{v_1^r}^2+\frac{7}{16}v_1^r v_2^r+\frac{81}{32}{v_2^r}^2\right)+v_1\ccdot v_2 v_2^2\left(\frac{7}{32}{v_1^r}^2+\frac{13}{16}v_1^r v_2^r+\frac{77}{32}{v_2^r}^2\right)\nonumber\\
&-&v_1\ccdot v_2\left(\frac{11}{64}{v_1^r}^4+\frac{5}{16}{v_1^r}^3v_2^r+\frac{15}{32}{v_1^r}^2{v_2^r}^2+\frac{11}{16}v_1^r{v_2^r}^3+\frac{65}{64}{v_2^r}^4\right)\nonumber\\
&+&v_1^2\left(\frac{27}{128}{v_1^r}^4+\frac{11}{32}{v_1^r}^3v_2^r+\frac{27}{64}{v_1^r}^2{v_2^r}^2+\frac{15}{32}v_1^r{v_2^r}^3+\frac{137}{128}{v_2^r}^4\right)\nonumber\\
&+&v_2^2\left(\frac{15}{128}{v_1^r}^4+\frac{9}{32}{v_1^r}^3v_2^r+\frac{33}{64}{v_1^r}^2{v_2^r}^2+\frac{27}{32}v_1^r{v_2^r}^3-\frac{27}{128}{v_2^r}^4\right)\nonumber\\
&-&\left.\frac{5}{128}{v_1^r}^6-\frac{5}{64}{v_1^r}^5v_2^r-\frac{15}{128}{v_1^r}^4{v_2^r}^2-\frac{5}{32}{v_1^r}^3{v_2^r}^3-\frac{25}{128}{v_1^r}^2{v_2^r}^4-\frac{15}{64}v_1^r{v_2^r}^5-\frac{5}{128}{v_2^r}^6\right]\vec{x}_1\nonumber\\
&+&\left[\frac{3}{16}\left(v_1\ccdot v_2\right)^2v_1^r-v_1^4\left(\frac{123}{64}v_1^r+\frac{49}{64}v_2^r\right)+v_1^2v_1\ccdot v_2\left(\frac{3}{16}v_1^r+\frac{1}{16}v_2^r\right)
-\frac{33}{32}v_1^2 v_2^2v_1^r\right.\nonumber\\
&+&v_1^2\left(\frac{77}{96}{v_1^r}^3+\frac{13}{32}{v_1^r}^2v_2^r+\frac{7}{32}v_1^r {v_2^r}^2+\frac{17}{96}{v_2^r}^3\right)
+v_1\ccdot v_2\left(\frac{1}{16}v_1^r+\frac{5}{16}v_2^r\right){v_1^r}^2\nonumber\\
&-&\left.\left.\left(\frac{13}{64}{v_1^r}^2+\frac{11}{64}v_1^r v_2^r+\frac{5}{32}{v_2^r}^2\right){v_1^r}^3\right]\vec{v}\right\}\nonumber\\
&+&\frac{G_N^2 m_1m_2}{r^2}\left\{m_2\left[\frac{5631}{320}v_1^4+\frac{2091}{80}\left(v_1\ccdot v_2\right)^2+\frac{9083}{960}v_2^4-\frac{10153}{240}v_1^2v_1\ccdot v_2
+\frac{9473}{480}v_1^2v_2^2-\frac{7193}{240}v_1\ccdot v_2 v_2^2\right.\right.\nonumber\\
&+&v_1^2\left(\frac{2}{5}v_1^r v_2^r-\frac{31}{5}{v_1^r}^2-\frac{239}{120}{v_2^r}^2\right)+v_2^2\left(\frac{241}{120}{v_2^r}^2+\frac{313}{120}v_1^r v_2^r-\frac{83}{15}{v_1^r}^2\right)+\frac{1133}{960}{v_1^r}^4\nonumber\\
&+&\left.v_1\ccdot v_2\left(\frac{247}{30}{v_1^r}^2+\frac{37}{60}v_1^r v_2^r+\frac{59}{60}{v_2^r}^2\right)
+\frac{1337}{240}{v_1^r}^3 v_2^r-\frac{7141}{480}{v_1^r}^2{v_2^r}^2+\frac{2197}{240}v_1^r{v_2^r}^3-\frac{4187}{960}{v_2^r}^4\right]\vec{x}_1\nonumber\\
&+&m_1\left[\frac{129}{80}\left(v_1\ccdot v_2\right)^2-\frac{1163}{960}v_1^4-\frac{2931}{320}v_2^4-\frac{7}{240}v_1^2v_1\ccdot v_2
-\frac{1973}{480}v_1^2v_2^2+\frac{2983}{240}v_1\ccdot v_2 v_2^2\right.\nonumber\\
&+&v_1^2\left(\frac{167}{60}{v_2^r}^2+\frac{1037}{120}v_1^r v_2^r-\frac{139}{30}{v_1^r}^2\right)+v_2^2\left(\frac{1409}{120}{v_1^r}^2-\frac{273}{20}v_1^r v_2^r+\frac{139}{20}{v_2^r}^2\right)+\frac{49}{320}{v_1^r}^4\nonumber\\
&-&\left.v_1\ccdot v_2\left(\frac{793}{120}{v_1^r}^2+\frac{247}{60}v_1^r v_2^r+\frac{329}{60}{v_2^r}^2\right)
+\frac{113}{240}{v_1^r}^3 v_2^r-\frac{4019}{480}{v_1^r}^2{v_2^r}^2+\frac{381}{80}v_1^r{v_2^r}^3-\frac{1133}{960}{v_2^r}^4\right]\vec{x}_1\nonumber\\
&+&m_1\left[v_1^2\left(\frac{9229}{480}v_1^r-\frac{8849}{480}v_2^r\right)+v_1\ccdot v_2\left(\frac{6499}{240}v_2^r-\frac{4529}{240}v_1^r\right)+v_2^2\left(\frac{2293}{160}v_1^r-\frac{3733}{160}v_2^r\right)\right.\nonumber\\
&-&\left.\left.\frac{3341}{480}{v_1^r}^3+\frac{10223}{480}{v_1^r}^2v_2^r-\frac{3781}{160}v_1^r{v_2^r}^2+\frac{4621}{480}{v_2^r}^3\right]\vec{v}\right\}+{\cal O}(G_N^3)\,.
\ee
As for the Energy, accelerations appearing at different PN orders have been 
systematically substituted via the equations of motion, truncated at the appropriate
PN order.

\section{Conclusions}
\label{se:conclusion}
We have computed the the conservative dynamics of a binary system within the
framework of the post-Newtonian approximation to general relativity at fourth post-Newtonian
order up to terms quadratic in the Newton constant.
By a systematic use of the effective field theory methods for non-relativistic
general relativity proposed by Goldberger and Rothstein it has been possible to
automatize the computation and derive the effective Lagrangian by summing the 
contributions of several Feynman diagrams.
We have also computed the energy function and verified Lorentz invariance,
giving a non trivial consistency check of our calculations.

This work is the first ingredient for the complete determination of the 4PN 
dynamics.

After the publication on the arXiv of the first version of this paper, \cite{Jaranowski:2012eb} 
appeared where the 4PN Hamiltonian has been computed in ADM coordinates in the center 
of mass frame, up to terms $O(G_N^2)$ and with the inclusion of some the $O(G_N^3)$ terms 
that could be determined by imposing Lorentz invariance.
A direct comparison of the two results can be made by computing the 
gauge invariant energy of circular orbits $E(x,\nu)$, depending only on the symmetric mass ratio
$\nu\equiv m_1m_2/m^2$, being $m=m_1+m_2$, and on the PN 
expansion parameter $x\equiv(G_N m \omega)^{2/3}$, with $\omega$ the rotation angular frequency.
We find perfect agreement with the new 4PN coefficients proportional to
$\nu^3$ and $\nu^4$ reported in \cite{Jaranowski:2012eb}.

\section*{Acknowledgments}
RS wishes to thank the theoretical physics department of the University of 
Geneva for kind hospitality and support during the preparation of part this 
work.
SF wishes to thank the physics department of the University of Urbino for 
hospitality during the preparation of part of this work.
The work of SF is supported by the FNS.

\end{document}